\documentclass[%
superscriptaddress,
aps,
prl,
twocolumn,
nolongbibliography
]{revtex4-2}

\usepackage{array}
\usepackage{dcolumn}
\usepackage{amsmath,amssymb,graphicx,comment,bm,color,dsfont,natbib,url}
\usepackage{tikz}
\begin{document}

\title{Semi-local one-dimensional optical wave turbulence}

\author{Cl\'ement Coll\'eaux}
\email{clement.colleaux@univ-cotedazur.fr}

\affiliation{Universit\'e C\^ote d'Azur, CNRS-Institut de Physique de Nice, 17 Rue Julien Laupr\^etre, Nice, 06200, France}

\author{Jonathan Skipp}
\email{j.skipp@aston.ac.uk}

\affiliation{Department of Applied AI and Robotics, College of Engineering and Physical Sciences, Aston University, Aston Triangle, Birmingham, B4 7ET, UK}
\affiliation{Aston Fluids Group, Aston University, Aston Triangle, Birmingham, B4 7ET, UK}

\author{Jason Laurie}
\email{j.laurie@aston.ac.uk}

\affiliation{Department of Mechatronics and Biomedical Engineering, College of Engineering and Physical Sciences, Aston University, Aston Triangle, Birmingham, B4 7ET, UK}
\affiliation{Aston Fluids Group, Aston University, Aston Triangle, Birmingham, B4 7ET, UK}

\author{Sergey Nazarenko}
\email{sergey.nazarenko@univ-cotedazur.fr}

\affiliation{Universit\'e C\^ote d'Azur, CNRS-Institut de Physique de Nice, 17 Rue Julien Laupr\^etre, Nice, 06200, France}

\date{\today}

\begin{abstract}
\noindent We study one-dimensional optical wave turbulence described by the 1D Schr{\"o}dinger–Helmholtz model for nonlinear light propagation in spatially nonlocal nonlinear optical media such as nematic liquid crystals. By exploiting the specific structure of the nonlocal response, we derive a reduced wave kinetic equation under a semi-local approximation that permits the study of weak-wave turbulent cascades. We explore the realisability of the wave turbulence predictions and demonstrate new turbulent behaviour related to spatial nonlocality. Moreover, we show strong dependence of the wave turbulence to the possible presence of incoherent solitonic structures.
\end{abstract}

\maketitle

\noindent \textbf{\textit{Introduction.}}
Light propagating through a nonlinear medium often enters a chaotic regime characterized by a broad spectrum of randomly interacting waves. Such states are called optical wave turbulence (OWT), with one-dimensional (1D) OWT realised in optical fibres~\cite{picozzi_spectral_2008, turitsyna_laminarturbulent_2013, suret_wave_2011} or liquid crystals~\cite{bortolozzo_optical_2009,laurie_one-dimensional_2012} being of particular interest due to their practical importance to optical telecommunications, and the possible presence of nonlinear structures like solitons~\cite{Colleaux2024boundstate}. Moreover, OWT is fundamentally interesting in its own right to study the mechanisms for energy transfer between scales, and the respective interaction of nonlinear structures among themselves, and with the underlying random wave background. A distinct feature of 1D optical systems is that in the first approximation, many are described by integrable nonlinear models within which the dynamics are represented by a set of coherent nonlinear solitons, freely passing through each other without distortion, embedded within a weak-wave field. Several works have explored integrable turbulence in the context of soliton gases~\cite{agafontsev_bound-state_2023,suret_soliton_2024}, rogue wave formation~\cite{soto-crespo_integrable_2016,randoux_nonlinear_2016,gelash_strongly_2018}, and wave field statistics~\cite{suret_wave_2011,randoux_intermittency_2014,roberti_early_2019,congy_statistics_2024}. However, integrability inhibits turbulent interactions and the formation of turbulent cascades described by the measurement of scale-invariant power-law spectra across scales. Thus, to describe such cascades, one has to consider deviations from integrability. Such examples arise naturally in the study of nonlinear light propagation in liquid crystals~\cite{bortolozzo_optical_2009, laurie_one-dimensional_2012}, optical fibres~\cite{picozzi_spectral_2008}, and optical systems with thermal nonlinearity~\cite{castillo_formation_1996, bekenstein_optical_2015,roger_optical_2016}.

In this Letter, we study 1D OWT in a non-integrable model with nonlocal nonlinearity, namely the 1D Schrödinger-Helmholtz equation (SHE). This model naturally arises in the study of nonlinear optics in liquid crystals~\cite{conti_route_2003, conti_observation_2004, bortolozzo_optical_2009}. By exploiting the structure of the nonlocal nonlinearity, we derive a reduced wave kinetic description using a semi-local approximation applicable to study scale-invariant turbulent cascades. Our 1D semi-local approximation model (SLAM) yields Kolmogorov-Zakharov power-law predictions for the stationary non-equilibrium evolution of the wave action spectrum in a weak turbulence regime. 
We theoretically study the reliability of our results and verify the SLAM predictions via direct numerical simulation.


\noindent \textbf{\textit{One-dimensional Schrödinger-Helmholtz equation.}} The 1D SHE is a nonlinear evolution equation for the complex scalar wavefunction  $\psi(x,t)$ in a periodic spatial domain $[0,L)$. The 1D SHE constitutes a non-integrable extension of the 1D nonlinear Schrödinger equation (NLSE) with a spatially nonlocal nonlinear potential $V(x,t)$:
\begin{align}\label{eq:she}
         i \frac{\partial \psi}{\partial t} =- \frac{1}{2} \frac{\partial^2 \psi}{\partial x^2} - V \psi,\qquad
         \Lambda V - \frac{\partial^2V}{\partial x^2} = \alpha |\psi|^2,
\end{align}
where the parameter $\Lambda$ controls the extent of the spatial nonlocality, while $\alpha$ is the normalised Kerr coefficient of the nonlinear interaction. We recover the 1D NLSE in the limit of $\alpha,\Lambda \to \infty$ with $\alpha/\Lambda$ constant. Equation~\eqref{eq:she} satisfies Hamilton's equation $i\partial \psi/\partial t = \delta H/\delta \psi^*$, with Hamiltonian 
\begin{align}\label{eq:ham}
H = \int \frac{1}{2}\left| \frac{\partial \psi}{\partial x} \right|^2 - \frac{\alpha}{2}\left[ \left(\Lambda -\frac{\partial^2}{\partial x^2}\right)^{-1/2}|\psi|^2\right]^2   \ {\rm d}x.
\end{align}
The first term on the right-hand side of~\eqref{eq:ham} is the quadratic energy associated with the free propagation of linear dispersive waves, while the second term is the energy contribution due to nonlinear interactions via the nonlocal potential $V$. In the absence of forcing and dissipation, the Hamiltonian $H$ is conserved under the evolution of Eq.~\eqref{eq:she}, as is the wave action 
\begin{align}\label{eq:waveaction}
N = \int |\psi|^2\ {\rm d}x.
\end{align}
Equation~\eqref{eq:she} can equivalently be written as a nonlinear evolution equation for the Fourier coefficient
$\hat{\psi}_{\bf k}(t) = \hat{\psi}({\bf k},t)=(1/L)\int_0^L \psi(x,t)e^{-i{\bf k}x}\ {\rm d}x$, where the physical-space solution is represented as 
$\psi(x,t) = \sum_{\bf k} \hat{\psi}_{\bf k}(t)e^{i{\bf k}x}$. (Note that we use the bold typeface for the signed 1D wavenumber ${\bf k}$, while $k=|{\bf k}|$.) 
This evolution equation in Fourier space is
\begin{align}\label{eq:she_fourier}
i\frac{\partial \hat{\psi}_{\bf k}}{\partial t} =\omega_{\bf k}\hat{\psi}_{\bf k} + \sum_{{\bf k}_1, {\bf k}_2, {\bf k}_3} T^{1,2}_{3,{\bf k}}\hat{\psi}_{1}\hat{\psi}_{2}\hat{\psi}_{3}^* \delta^{1,2}_{3,{\bf k}},
\end{align}
where $\omega_{\bf k}=\omega({\bf k})=k^2/2$ is the linear frequency dispersion relation, $\hat{\psi}_{j} = \hat{\psi}_{{\bf k}_j}$ for $j=1,2,3$, the four-wave interaction coefficient $T^{1,2}_{3,{\bf k}} =T^{{\bf k}_1,{\bf k}_2}_{{\bf k}_3,{\bf k}} =  T({\bf k}_1,{\bf k}_2,{\bf k}_3,{\bf k})$ is given by
\begin{align}\label{eq:T}
T^{1,2}_{3,{\bf k}} = -\frac{\alpha}{4}&\left[\frac{1}{({\bf k}_1-{\bf k})^2+\Lambda} +\frac{1}{({\bf k}_2-{\bf k})^2+\Lambda}\right.\nonumber\\
&\left.+\frac{1}{({\bf k}_1-{\bf k}_3)^2+\Lambda} +\frac{1}{({\bf k}_2-{\bf k}_3)^2+\Lambda} \right],
\end{align}
and $\delta^{1,2}_{3,{\bf k}} = \delta({\bf k}_1+{\bf k}_2-{\bf k}_3-{\bf k})$ is a Kronecker delta  involving the difference of wavenumbers. Equation~\eqref{eq:she_fourier} describes the propagation of dispersive waves with a dispersion relation $\omega_{\bf k}$ that undergo four-wave mixing of the $2\leftrightarrow 2$ type, whose interaction kernel is $T^{1,2}_{3,{\bf k}}$.

The main focus of the wave turbulence approach is to describe the evolution of the wave action spectrum $n_{\bf k}=(L/2\pi)\langle |\hat{\psi}_{\bf k}|^2\rangle$ where $\langle \cdot\rangle$ denotes ensemble averaging over random initial conditions. The derivation of an integro-differential wave kinetic equation (WKE) for the wave action spectrum $n_{\bf k}$ follows a standard procedure outlined in Refs.~\cite{zakharov_kolmogorov_1992,nazarenko_wave_2011}. 

The linear dispersion relation $\omega_{\bf k}\propto k^\beta$ having an exponent $\beta>1$, together with the single spatial dimension, mean that the four-wave $2\leftrightarrow 2$ processes governed by the nonlinear term in Eq.~\eqref{eq:she_fourier} do not contribute to resonant four-wave mixing at leading order. 
Consequently, one must perform a quasi-linear canonical transformation to a new Fourier variable to remove leading non-resonant wave interactions, until the leading wave mixing process is resonant. This procedure has been applied in many wave turbulence contexts, including water waves~\cite{krasitskii_reduced_1994,dyachenko_five-wave_1995}, superfluid Kelvin waves~\cite{kozik_kelvin-wave_2004,laurie_interaction_2010}, and nonlinear optics~\cite{bortolozzo_optical_2009,laurie_one-dimensional_2012}. In the case of the 1D SHE~\eqref{eq:she}, the leading resonant wave process becomes a six-wave $3\leftrightarrow 3$ resonant wave interaction, leading to the following WKE,
\begin{align}\label{eq:kinetic}
&\frac{\partial n_{\bf k}}{\partial t} = 24 \pi \! \!\int \left|W^{1,2,3}_{4,5,{\bf k}}\right|^2 \!
\left[ \frac{1}{n_{{\bf k}}} +\frac{1}{n_5}+\frac{1}{n_4}-\frac{1}{n_1}-\frac{1}{n_2}-\frac{1}{n_3} \right]\nonumber\\
&\times \!  n_1n_2n_3n_4n_5n_{\bf k}\, \delta^{1,2,3}_{4,5,{\bf k}} \,\delta(\omega^{1,2,3}_{4,5,{\bf k}})\, {\rm d}{{\bf k}}_1{\rm d}{\bf k}_2{\rm d}{{\bf k}}_3{\rm d}{\bf k}_4{\rm d}{\bf k}_5.
\end{align}
The right-hand side of Eq.~\eqref{eq:kinetic} is called the collision integral and is a five-dimensional integral over wavenumber space. Here, $\delta^{1,2,3}_{4,5,{\bf k}}=\delta({\bf k}_1+{\bf k}_2+{\bf k}_3-{\bf k}_4-{\bf k}_5-{\bf k})$ and $\delta(\omega^{1,2,3}_{4,5,{\bf k}})$ are Dirac delta functions that constrain the wave sextets to the resonant manifold defined by ${\bf k}_1+{\bf k}_2+{\bf k}_3-{\bf k}_4-{\bf k}_5-{\bf k}=0$ and $\omega^{1,2,3}_{4,5,{\bf k}}:=\omega_1+\omega_2+\omega_3-\omega_4-\omega_5-\omega_{\bf k}=0$. 
The six-wave interaction coefficient $W^{1,2,3}_{4,5,{\bf k}}$ arises as a consequence of the canonical transformation, and describes the coupling of two non-resonant four-wave interaction processes to form a resonant six-wave process of the form $3\leftrightarrow 3$. The explicit derivation and the final expression for $W^{1,2,3}_{4,5,{\bf k}}$ are given in the Supplemental Material (SM).

 WKEs describing $n\leftrightarrow n$ wave interactions for $n\in \mathbb{N}$, such as Eq.~\eqref{eq:kinetic},  conserve two quadratic invariants, namely the wave energy $E$ and the wave action $N$:
\begin{align}\label{eq:invariants}
E = \int \omega_{\bf k} n_{\bf k} \ {\rm d} {\bf k}, \quad \text{and}\quad N = \int n_{\bf k}\  {\rm d}{\bf k},
\end{align}
which are Fourier representations of the quadratic energy in Eq.~\eqref{eq:ham}, and of the wave action Eq.~\eqref{eq:waveaction}, respectively.
Evolution under the  WKE redistributes the spectral densities of $E$ and $N$ across wavenumber space, in a way predicted by the dual-cascade argument of Fjørtoft~\cite{fjortoft_changes_1953}. This argument is recapitulated in the context of wave turbulence in many places, see Book~\cite{nazarenko_wave_2011}. It concludes that due to the the relationship between the spectral densities of $E$ and $N$, energy must predominately evolve towards small scales and wave action to large scales.

With the addition of spectrally localised external forcing and dissipation, the WKE admits 
power-law solutions on which either of the quadratic invariants $E$ or $N$ are transported with constant flux 
across an inertial range of scales. On these solutions, called Kolmogorov-Zakharov (KZ) spectra, the collision integral of~\eqref{eq:kinetic} integrates to zero, implying statistical stationarity $\partial n_{\bf k}/\partial t=0$.  
Traditionally, to determine the KZ spectra, one uses scale invariance of the WKE to make so-called Zakharov transforms, which
map different integration regions of the collision integral onto one another. This allows one to formally derive the KZ power-law exponents upon which the collision integral vanishes~\cite{zakharov_kolmogorov_1992}. However, the lack of scale invariance of Eq.~\eqref{eq:T}, results in the six-wave interaction coefficient $W^{1,2,3}_{4,5,{\bf k}}$, and hence the WKE~\eqref{eq:kinetic}, not possessing scale invariance.
Hence, the Zakharov transform cannot be applied, with the KZ power-law solutions determined only for specific limiting cases of the SHE when $\Lambda\to 0$, or $\Lambda\to \infty$~\cite{skipp_wave_2020}. Moreover, the limiting case $\Lambda\to 0$ is delicate because the WKE collision integral in~\eqref{eq:kinetic} is divergent for $\Lambda=0$ (cf. Ref.~\cite{skipp_effective_2023}). Thus, a careful treatment of the $\Lambda\to 0$ limit is needed. This is the focus of the present work.

\noindent \textbf{\textit{The semi-local approximation.}}
The SHE has a particular structure where, in the limit of $\Lambda\to 0$, the nonlinear interaction coefficient~\eqref{eq:T} is locally peaked for pairs of wavenumbers close to each other, e.g. such pairing as ${\bf k}_1 \approx {\bf k}_3$ and ${\bf k}_2 \approx {\bf k}$. By considering contributions of this kind, we can apply a semi-local approximation to the WKE and derive an effective semi-local approximation model (SLAM) for studying nonlocal 1D OWT.

Derivation of the SLAM closely follows the strategy outlined in Ref.~\cite{skipp_effective_2023} where the 2D SHE was studied. However, in the present case, the WKE is of six-wave type, and hence, the approach differs and is distinct. We consider the dominant contribution to the WKE when $\Lambda \ll k^2$, where $k$ represents the characteristic wavenumber magnitude being considered. Then $W^{1,2,3}_{4,5,{\bf k}}$ becomes sharply peaked when pairs of wavenumbers become comparable. Indeed, to satisfy the six-wave resonant manifold conditions, we find, without loss of generality, that as two wavenumbers become comparable, i.e. ${\bf k}_1\approx {\bf k}_4$, then we require additionally that ${\bf k}_2 \approx {\bf k}_5$ and ${\bf k}_3\approx {\bf k}$. It is precisely this behaviour that we leverage when deriving our SLAM. Specific details are left to the SM. However, we summarise our approach as follows: we denote two small variables ${\bf p}={\bf k}_1-{\bf k}_4$ and ${\bf q}={\bf k}_2-{\bf k}_5$ and perform an expansion of the collision integral in the semi-local limit of $p \ll {k}_1, {k}_4$; $q \ll {k}_2, {k}_5$, and $\Lambda \ll k^2$.  By application of a test function, we simplify $W^{1,2,3}_{4,5,{\bf k}}$ in the semi-local limit, and integrate out the two resonant manifold Dirac delta functions to eventually reduce the dimensionality of the collision integral by three. With a final integration by parts, we simplify the WKE into a continuity equation for the wave action flux $Q$, which we call the 1D SLAM:
\begin{widetext}

\vspace{-15pt}

\begin{align}\label{eq:slam}
\frac{\partial n_{\bf k}}{\partial t} &= - \frac{\partial Q}{\partial {\bf k}}, 
\quad
Q({\bf k}) = \! -\frac{1}{\Lambda^{5/2}}\! \! \int  \! \! V^{1,2}_{\bf k} \!  \left({\bf k}_2-{\bf k}_1\right) \! \left[ ({\bf k}-{\bf k}_2)n_{{\bf k}}^2n_{2}^2\frac{\partial n_{1}}{\partial {\bf k}_1}
+({\bf k}_1-{\bf k})n_{{\bf k}}^2n_{1}^2\frac{\partial n_{2}}{\partial {\bf k}_2}
+({\bf k}_2-{\bf k}_1)n_{1}^2n_{2}^2\frac{\partial n_{{\bf k}}}{\partial {\bf k}} \right]\! {\rm d}{\bf k}_1 {\rm d} {\bf k}_2.
\end{align}

\vspace{-10pt}

\end{widetext}
Here, the effective interaction coefficient $V^{1,2}_{\bf k}$ is given by Eq.~\eqref{eq:V12k},
which is obtained from the original coefficient $W^{1,2,3}_{4,5,{\bf k}}$ under the semi-local approximation, and is explicitly derived in the SM. Incidentally, the expansions we use in the derivation fail around the singular point where ${\bf k}_1$ and ${\bf k}_2$ tend to ${\bf k}$ simultaneously. This results in a spurious divergence of the integral defining $Q$ at this point, even though the original collision integral is convergent.  In the SM, we remedy this problem by modifying $V^{1,2}_{\bf k}$ using an idea inspired by the theory of sticky particles -- the so-called ``collisional efficiency'' regularisation~\cite{horvai_coalescence_2008, pearson1984monte}, see Eq.~\eqref{eq:VR}.
Importantly, $V^{1,2}_{\bf k}$ has the symmetries $V^{1,2}_{\bf k}=V^{2,1}_{\bf k}=V^{1,{\bf k}}_2$. It also possesses scale invariance $V^{\lambda{\bf k}_1,\lambda{\bf k}_2}_{\lambda{\bf k}} = |\lambda|^{-7}V^{{\bf k}_1,{\bf k}_2}_{\bf k}$. This means that the 1D SLAM~\eqref{eq:slam} is scale-invariant and amenable to the standard approaches of wave turbulence theory. Moreover, the original quadratic invariants $N$ and $E$ remain conserved by the 1D SLAM (see the SM).

If the wave action spectrum is symmetric with respect to the sign of ${\bf k}$, then $n_{\bf k} = n(k)$, i.e.\ it becomes a function of the wavenumber magnitude $k$ only, and the continuity equation can be represented as 
$\partial n_{\bf k}/\partial t = -\partial Q_s/\partial k$ where $Q_s(k) = {\rm sgn}({\bf k})Q({\bf k})$ with the understanding that $k>0$.

\noindent\textbf{\textit{Stationary solutions.}} The study of stationary $(\partial n_{\bf k}/\partial t =0)$ solutions of WKEs, such as Eq.~\eqref{eq:kinetic}, is central to the predictions of 
wave turbulence. The 1D SLAM~\eqref{eq:slam} has a stationary solution of thermal equilibrium 
\begin{align}\label{eq:rayleigh-jeans}
n_{\bf k} = \frac{T}{\mu + \omega_{\bf k}},
\end{align}
with constants being the temperature $T$ and the chemical potential $\mu$.
Spectrum~\eqref{eq:rayleigh-jeans} is known as the 
Rayleigh-Jeans spectrum. It admits zero flux, resulting in the vanishing of the integral in~\eqref{eq:slam}. 

The scale invariance of the 1D SLAM potentially permits another class of stationary solutions to exist. These are the KZ spectra, which describe a constant flux transfer of invariants via a self-similar, scale-by-scale cascade through an inertial range of scales. 
For them to be realised, one must add external narrow-band forcing and dissipation to the right-hand side of Eq.~\eqref{eq:slam}.
As our system has two quadratic invariants: $E$ and $N$, we expect a dual cascade, analogous to 2D turbulence~\cite{boffetta_two-dimensional_2012}, where each cascade corresponds to the local transfer of one of the invariants.
The waveaction cascade is inverse, meaning that the flow of $N$ is from small to large scales, while the energy cascade is direct, meaning that $E$ flows from large to small scales, as predicted by the Fj{\o}rtoft argument~\cite{fjortoft_changes_1953}.

By considering a self-similar power-law solution of the form $n_{\bf k} = C k^{x}$, we can express the wave action flux in dimensionless form $Q_s(k) =  k^{5x-4}I(x)$ with 
\begin{align}\label{eq:I}
I(x) &= -\frac{C^5}{\Lambda^{5/2}}\! \int x V^{{\bf s}_1,{\bf s}_2}_{1}\left({\bf s}_2-{\bf s}_1\right)\! \left[ {\rm sgn}({\bf s}_1)(1-{\bf s}_2)s_2^{2x}s_1^{x-1}\right.\nonumber\\ &\left.+{\rm sgn}({\bf s}_2)({\bf s}_1-1)s_1^{2x}s_2^{x-1} +({\bf s}_2-{\bf s}_1)s_1^{2x}s_2^{2x} \right] \! {\rm d}{\bf s}_1 {\rm d}{\bf s}_2,
\end{align}
where ${\bf s}_1 = {\bf k}_1/k$ and ${\bf s}_2 = {\bf k}_2/k$. Observe that $x=4/5$ provides a $k$-independent expression for the wave action flux, which is finite if the non-dimensional integral $I(4/5)$ is convergent. Consequently, the KZ solution $n_{\bf k} = Ck^{4/5}$ supports a non-equilibrium steady state characterised by a constant negative wave action flux $Q_s$, indicating an inverse cascade. 
By contrast, the corresponding energy flux $P_s$ will necessarily vanish. The energy flux $P_s$ can be determined via integration by parts $P_s(k) = k^2Q_s/2 - \int_0^k k'Q_s(k')\, \mathrm{d}k'$, leading to
\begin{align}\label{eq:energyflux}
P_s(k) = \frac{k^{5x-2}(5x-4)}{10x-4}I(x).
\end{align}
Notice that on the inverse wave action spectrum $x=4/5$, we have $P_s=0$, assuming again that $I(4/5)$ remains finite. We can study the convergence of the integral $I(x)$ with respect to $x$ by carefully considering the limiting behaviour when, either or both, ${\bf s}_1, {\bf s}_2 \to -\infty, 0$, or  $\infty$. Details of this are presented in the SM. We find that $I(x)$ is convergent only for $0\leq x < 3/4$. 
This means that $I(4/5)$ results in a divergent integral. Therefore, an inverse wave action KZ spectrum is unrealisable by the dynamics. Physically, this implies that the wave interactions are dominated by nonlocal contributions, which turn out from our analysis to be dominant when $s_1, s_2 \to \infty$ simultaneously.  Consequently, additional analysis based on the leading nonlocal contribution is required to examine this solution. We will return to this in the following section.

The second KZ solution describes a constant positive flux of wave energy, i.e.\ $P_s > 0$. The prediction of its power-law index can be ascertained from the corresponding energy flux equation $\partial (\omega_{\bf k} n_{\bf k})/\partial t = -\partial P_s/\partial k$ with $P_s$ given as in Eq.~\eqref{eq:energyflux}. Observe that a spectral slope of $x=2/5$ implies a $k-$independent wave energy flux; however, it results in $P_s = -2 I(2/5)/0$, which is divergent unless $I(2/5)$ vanishes. We show in the SM that $I(2/5)$ is equal to zero after applying the Zakharov transformation and is consistent with the convergence criterion for $I(x)$ we stated earlier. The value of the wave energy flux can be determined via L'H\^opital's rule, giving $P_s = -I'(2/5)/5$. Moreover, when $x=2/5$, the wave action flux $Q_s$ vanishes, indicating that this is a valid KZ wave energy cascade solution of the 1D SLAM.

\noindent\textbf{\textit{Nonlocal analysis for the inverse wave action KZ solution.}}
Returning to the first KZ solution $x=4/5$ for the inverse wave action cascade, recall that the integral $I(4/5)$ is divergent in the simultaneous limit $s_1, s_2 \to \infty$. One can perform a nonlocal analysis of this KZ solution in the limit of the leading divergence; details are presented in the SM. We consider the limit of Eq.~\eqref{eq:slam} when $k\ll k_1,k_2$, simplifying the interaction coefficient accordingly. Then the wave action flux $Q_s$ given in Eq.~\eqref{eq:slam} is reduced to 
\begin{align}\label{eq:Qnonlocal}
Q_s = -D \frac{\partial n_{\bf k}}{\partial k},
\end{align}
where $D=\Lambda^{-5/2} \int V^{{\bf k}_1,{\bf k}_2}_0 
({\bf k}_1 -{\bf k}_2)^2
n_{1}^2 n_2^2\ {\rm d}{\bf k}_1 {\rm d}{\bf k}_2$. This leads to a diffusion equation 
\begin{align}\label{eq:diffusion}
    \frac{\partial n_{\bf k}}{\partial t} = D\frac{\partial^2 n_{\bf k}}{\partial k^2}.
\end{align}
Consequently, a steady state power-law solution of~\eqref{eq:diffusion} leads to a \textit{nonlocal} wave action spectrum prediction of the form
$n_{\bf k} \propto k$. Since $D>0$, this spectrum corresponds to an inverse wave action cascade, $Q_s={\rm const.} <0$. Note that the linear exponent ($x=1$) is consistent with our assumption of nonlocality, as the integral $I(1)$ is divergent precisely in the nonlocal limit of $s_1,s_2\to \infty$ simultaneously. This divergence is regularized in the integral defining $D$ because $n_{\bf k}$ is cut off at the ultra-violet end of Fourier space by the dissipation.

\noindent \textbf{\textit{Numerical simulations.}}
We solve the 1D SHE~\eqref{eq:she} using a spatial Fourier pseudo-spectral method in a periodic domain $[0,L)$ of length $L=2\pi$ consisting of $N=4096$ uniform grid points. The nonlinear term is computed in physical space using a $3/2-$dealiasing rule. Time integration is performed using a fourth-order exponential time-differencing Runge-Kutta method with a fixed time step of $dt = 1\times 10^{-6}$. The SHE parameters in all of the simulations are $\alpha=1$ and $\Lambda=1$, which is consistent with the limit in which we derive the 1D SLAM~\eqref{eq:slam}. 
As we investigate statistically stationary, forced-dissipated wave turbulence, we include additive forcing $+if(x,t)$ and localised dissipation $-id(x,t)$ to the right-hand side of the first equation in~\eqref{eq:she}. The forcing is localised in Fourier space around a wavenumber $k_f$. We define it as $f(x,t) = \sum_{\bf k} \hat{f}_{\bf k}\eta_{\bf k}(t)e^{i{\bf k}x}$, with amplitudes $\hat{f}_{\bf k} = {\rm const.}$ for $|k-k_f|<\Delta k$ and $0$ otherwise, with $\eta_{\bf k}(t)$ being independent standard complex normal random variables for each wavenumber ${\bf k}$. The dissipation is of the form of hypo- and hyper-viscosity localised towards either end of wavenumber space with $d(x,t) = \gamma (-\partial^2/\partial x^2)^{-4}\psi + \nu (-\partial^2/\partial x^2)^{2}\psi$ with coefficients $\gamma$ and $\nu$.

\begin{figure}[t]
   \begin{center}
    \includegraphics[width=\linewidth]{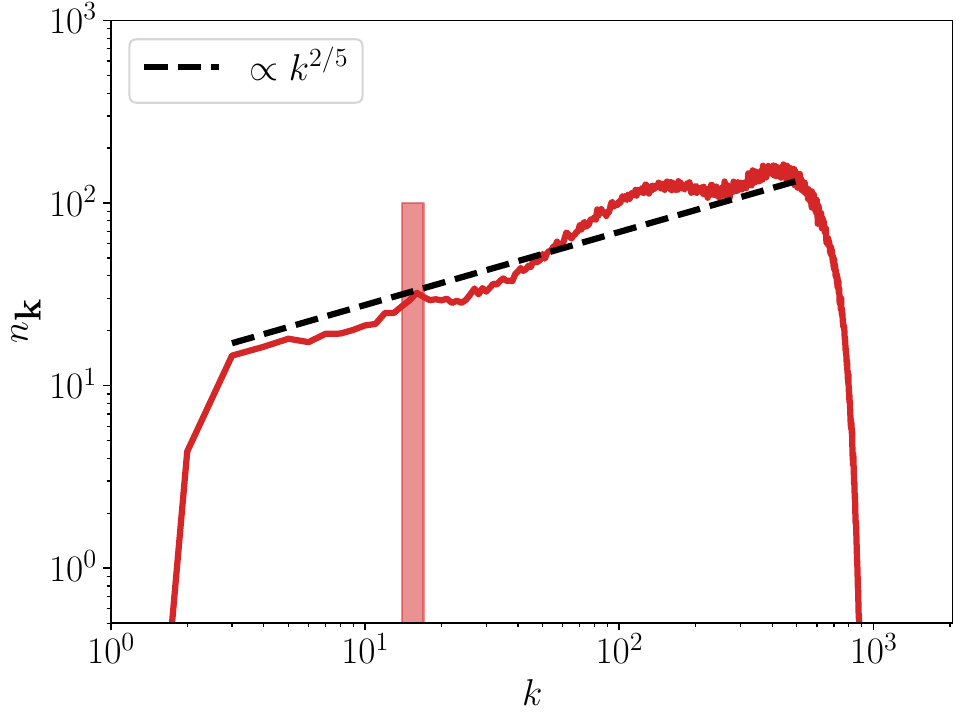}
    \caption{Wave action spectrum $n_{\bf k}$ for the direct energy cascade. The additive forcing with $k_f=16$ and $\Delta k=2$ (light red rectangle). The KZ power-law scaling $n_{\bf k}\propto k^{2/5}$ is shown by the black dashed line.\label{fig:direct}}
    \end{center}
\end{figure}

We numerically study the direct energy cascade by situating the forcing at $k_f=16$ with $\Delta k=2$, with forcing amplitude $\hat{f}_{\bf k}=100$. The dissipation coefficients are $\gamma=2\times 10^{6}$ and $\nu = 1\times 10^{-46}$, which ensures that the dissipation acts locally at the extreme ends of Fourier space. We evolve the system from a zero-state and wait until the statistical stationarity of the wave action $N$ and Hamiltonian $H$ is observed. Only then do we perform the analysis. The wave action spectrum $n_{\bf k}$ is presented in Fig.~\ref{fig:direct} and is averaged over a long time window during statistical steady state conditions. We observe that the wave action spectrum does not display a clear power-law scaling across the whole inertial range. Rather, a steeper than KZ scaling from the forcing scale $k_f$ to about $k\simeq 100$ is observed, before a series of gentle undulations until the dissipation region. However, the average slope of the spectrum is close to the KZ prediction of $n_{\bf k}\propto k^{-2/5}$, which we expect to observe.  

\begin{figure}[t]
   \begin{center}
    \includegraphics[width=\linewidth]{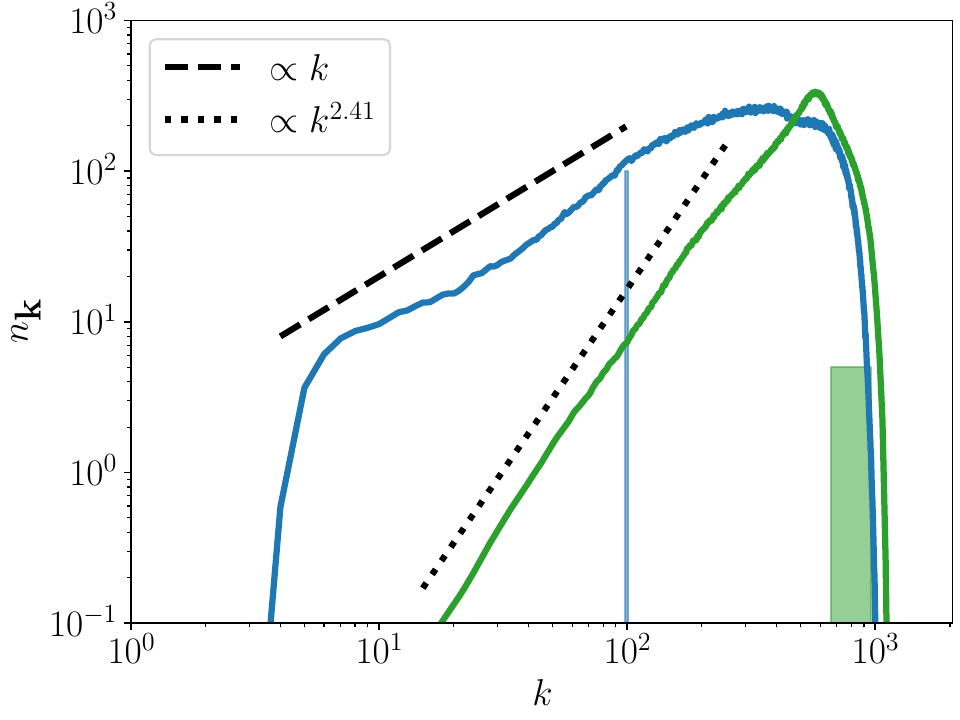}
    \caption{Wave action spectrum $n_{\bf k}$ for the inverse wave action cascade, for two different forcings. The blue curve corresponds to additive forcing with $k_f=100$ and $\Delta k=2$ (light blue rectangle). The green curve is an additional run with a broader smaller-scale forcing of $k_f=812$ and $\Delta k=150$ (light green rectangle). The nonlocal power-law scaling $n_{\bf k}\propto k$ is shown by the black dashed line, while a steeper, numerically fitted, spectrum of $n_{\bf k}\propto k^{2.41}$ is given by the black dotted line.\label{fig:inverse}}
    \end{center}
\end{figure}

An initial simulation of the inverse wave action cascade was performed with forcing at $k_f=100$ with $\Delta k=2$, and forcing amplitude $\hat{f}_{\bf k}=120$. In this case, the dissipation coefficients are $\gamma=2\times 10^{9}$ and $\nu = 1\times 10^{-36}$. The wave action spectrum $n_{\bf k}$ is presented in Fig.~\ref{fig:inverse} as the blue curve.
We observe that the wave action spectrum is consistent with our nonlocal theory prediction of $n_{\bf k}\propto k$ from the forcing scale $k_f$ down to $k\simeq 10$ where we observe a gradual plateauing of the spectrum before the hypo-viscosity dissipates the spectrum. It is clear that the formal KZ solution of $n_{\bf k}\propto k^{4/5}$ is not realised, as the locality analysis suggests. 

\begin{figure}[t]
    \begin{center}
    \includegraphics[width =\linewidth]{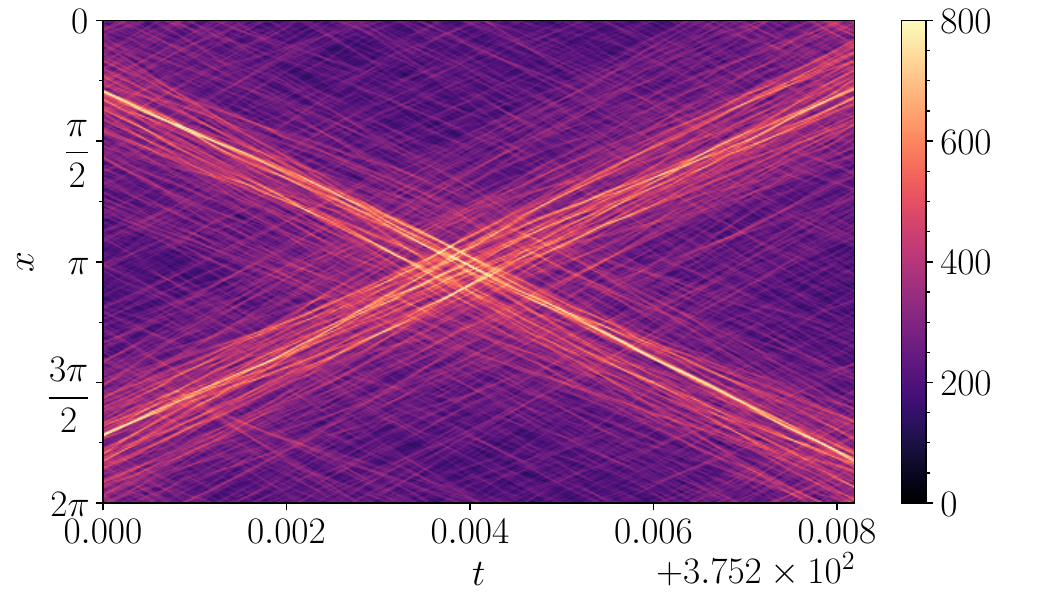}
    \caption{Spatial-temporal plot of the intensity $|\psi(x,t)|^2$ from $t=375.2$ for the inverse wave action cascade simulation corresponding to the green curve in Fig.~\ref{fig:inverse}. \label{fig:inverse-xt}}
    \end{center}
\end{figure}

We find that the inverse wave action spectrum is sensitive to the position and strength of the additive forcing. For example, in Fig.~\ref{fig:inverse} we present a second simulation (green curve) with a weaker but broader forcing spectrum of $k_f=812$, $\Delta k=150$, and $\hat{f}_{\bf k} = 5$. We see a much steeper power-law spectrum than before, with an exponent close to $x=2.41$, which is not theoretically predicted. Inspection of the physical space state, presented in Fig.~\ref{fig:inverse-xt}, reveals the appearance of two oppositely propagating localised structures of waves, reminiscent of incoherent solitons observed in a variety of nonlocal NLSEs~\cite{picozzi_incoherent_2011,laurie_one-dimensional_2012,garnier_incoherent_2021}. As Fig.~\ref{fig:inverse-xt} displays, our structures are composed of many fine filaments that weave between each other but collectively propagate in straight lines with a speed close to the group velocity $v_g=\partial \omega_{\bf k}/\partial {k} = k\simeq 573$ of linear waves at the scale where the wave action spectrum is at its maximum. The presence of such structures invalidates the homogeneity assumption of weak-wave turbulence approach and the KZ prediction.

\textbf{\textit{ Conclusions.}}
In this Letter, we derived a new reduced model of 1D optical wave turbulence in nonlocal media, the SLAM, and found power-law scaling solutions corresponding to the direct cascade of energy and the inverse cascade of wave action. The direct cascade is shown to be formed by local wave interactions, whereas the inverse cascade is dominated by the nonlocal interactions in the limit of two small-scale waves. We performed direct numerical simulations of the original SHE model and confirmed the theoretical predictions from SLAM in the weakly forced and dissipated regime. For stronger forcing, we have observed incoherent solitonic structures that break the homogeneity of the system and drastically change the scaling properties of the wave turbulence. We emphasise that a complete theory of wave turbulence would include a unified description of solitonic structures together with weakly nonlinear waves. The development of such a unified theory is an important task for future research.

Beyond the immediate application to optical systems, this work outlines a methodology for studying wave turbulence in systems with spatially nonlocal interactions. In the case examined in this Letter, the nonlocality is spectrally peaked in Fourier space, which selects a narrow band of wave interactions. This semilocal property allows the wave kinetics of the system to be reduced to a SLAM.
We envisage that this method may be applicable beyond the realm of nonlinear optics, to other nonlocal systems. For example, Kelvin waves on quantum vortex lines~\cite{boffetta_modeling_2009,laurie_interaction_2010} in the Gross-Pitaevskii equation where a spatial nonlocality is introduced to model the so-called roton minimum~\cite{muller_kolmogorov_2020}.

\textbf{\textit{ Acknowledgments.}}
\noindent This work has been supported by the European Union's Horizon2020 Marie Skłodowska-Curie actions through the research innovation and staff-exchange grant HALT (no. 823937), and by the Simons Foundation Collaboration grant Wave Turbulence (Award ID 651471). J.L. and J.S. are supported by the Leverhulme Trust project grant no. RPG-2021-014.

\bibliography{library}

\begin{thebibliography}{36}%
\makeatletter
\providecommand \@ifxundefined [1]{%
 \@ifx{#1\undefined}
}%
\providecommand \@ifnum [1]{%
 \ifnum #1\expandafter \@firstoftwo
 \else \expandafter \@secondoftwo
 \fi
}%
\providecommand \@ifx [1]{%
 \ifx #1\expandafter \@firstoftwo
 \else \expandafter \@secondoftwo
 \fi
}%
\providecommand \natexlab [1]{#1}%
\providecommand \enquote  [1]{``#1''}%
\providecommand \bibnamefont  [1]{#1}%
\providecommand \bibfnamefont [1]{#1}%
\providecommand \citenamefont [1]{#1}%
\providecommand \href@noop [0]{\@secondoftwo}%
\providecommand \href [0]{\begingroup \@sanitize@url \@href}%
\providecommand \@href[1]{\@@startlink{#1}\@@href}%
\providecommand \@@href[1]{\endgroup#1\@@endlink}%
\providecommand \@sanitize@url [0]{\catcode `\\12\catcode `\$12\catcode `\&12\catcode `\#12\catcode `\^12\catcode `\_12\catcode `\%12\relax}%
\providecommand \@@startlink[1]{}%
\providecommand \@@endlink[0]{}%
\providecommand \url  [0]{\begingroup\@sanitize@url \@url }%
\providecommand \@url [1]{\endgroup\@href {#1}{\urlprefix }}%
\providecommand \urlprefix  [0]{URL }%
\providecommand \Eprint [0]{\href }%
\providecommand \doibase [0]{https://doi.org/}%
\providecommand \selectlanguage [0]{\@gobble}%
\providecommand \bibinfo  [0]{\@secondoftwo}%
\providecommand \bibfield  [0]{\@secondoftwo}%
\providecommand \translation [1]{[#1]}%
\providecommand \BibitemOpen [0]{}%
\providecommand \bibitemStop [0]{}%
\providecommand \bibitemNoStop [0]{.\EOS\space}%
\providecommand \EOS [0]{\spacefactor3000\relax}%
\providecommand \BibitemShut  [1]{\csname bibitem#1\endcsname}%
\let\auto@bib@innerbib\@empty
\bibitem [{\citenamefont {Picozzi}\ \emph {et~al.}(2008)\citenamefont {Picozzi}, \citenamefont {Pitois},\ and\ \citenamefont {Millot}}]{picozzi_spectral_2008}%
  \BibitemOpen
  \bibfield  {author} {\bibinfo {author} {\bibfnamefont {A.}~\bibnamefont {Picozzi}}, \bibinfo {author} {\bibfnamefont {S.}~\bibnamefont {Pitois}},\ and\ \bibinfo {author} {\bibfnamefont {G.}~\bibnamefont {Millot}},\ }\href {https://doi.org/10.1103/PhysRevLett.101.093901} {\bibfield  {journal} {\bibinfo  {journal} {Physical Review Letters}\ }\textbf {\bibinfo {volume} {101}},\ \bibinfo {pages} {093901} (\bibinfo {year} {2008})},\ \bibinfo {note} {publisher: American Physical Society}\BibitemShut {NoStop}%
\bibitem [{\citenamefont {Turitsyna}\ \emph {et~al.}(2013)\citenamefont {Turitsyna}, \citenamefont {Smirnov}, \citenamefont {Sugavanam}, \citenamefont {Tarasov}, \citenamefont {Shu}, \citenamefont {Babin}, \citenamefont {Podivilov}, \citenamefont {Churkin}, \citenamefont {Falkovich},\ and\ \citenamefont {Turitsyn}}]{turitsyna_laminarturbulent_2013}%
  \BibitemOpen
  \bibfield  {author} {\bibinfo {author} {\bibfnamefont {E.~G.}\ \bibnamefont {Turitsyna}}, \bibinfo {author} {\bibfnamefont {S.~V.}\ \bibnamefont {Smirnov}}, \bibinfo {author} {\bibfnamefont {S.}~\bibnamefont {Sugavanam}}, \bibinfo {author} {\bibfnamefont {N.}~\bibnamefont {Tarasov}}, \bibinfo {author} {\bibfnamefont {X.}~\bibnamefont {Shu}}, \bibinfo {author} {\bibfnamefont {S.~A.}\ \bibnamefont {Babin}}, \bibinfo {author} {\bibfnamefont {E.~V.}\ \bibnamefont {Podivilov}}, \bibinfo {author} {\bibfnamefont {D.~V.}\ \bibnamefont {Churkin}}, \bibinfo {author} {\bibfnamefont {G.}~\bibnamefont {Falkovich}},\ and\ \bibinfo {author} {\bibfnamefont {S.~K.}\ \bibnamefont {Turitsyn}},\ }\href {https://doi.org/10.1038/nphoton.2013.246} {\bibfield  {journal} {\bibinfo  {journal} {Nature Photonics}\ }\textbf {\bibinfo {volume} {7}},\ \bibinfo {pages} {783} (\bibinfo {year} {2013})}\BibitemShut {NoStop}%
\bibitem [{\citenamefont {Suret}\ \emph {et~al.}(2011)\citenamefont {Suret}, \citenamefont {Picozzi},\ and\ \citenamefont {Randoux}}]{suret_wave_2011}%
  \BibitemOpen
  \bibfield  {author} {\bibinfo {author} {\bibfnamefont {P.}~\bibnamefont {Suret}}, \bibinfo {author} {\bibfnamefont {A.}~\bibnamefont {Picozzi}},\ and\ \bibinfo {author} {\bibfnamefont {S.}~\bibnamefont {Randoux}},\ }\href {https://doi.org/10.1364/OE.19.017852} {\bibfield  {journal} {\bibinfo  {journal} {Optics Express}\ }\textbf {\bibinfo {volume} {19}},\ \bibinfo {pages} {17852} (\bibinfo {year} {2011})}\BibitemShut {NoStop}%
\bibitem [{\citenamefont {Bortolozzo}\ \emph {et~al.}(2009)\citenamefont {Bortolozzo}, \citenamefont {Laurie}, \citenamefont {Nazarenko},\ and\ \citenamefont {Residori}}]{bortolozzo_optical_2009}%
  \BibitemOpen
  \bibfield  {author} {\bibinfo {author} {\bibfnamefont {U.}~\bibnamefont {Bortolozzo}}, \bibinfo {author} {\bibfnamefont {J.}~\bibnamefont {Laurie}}, \bibinfo {author} {\bibfnamefont {S.}~\bibnamefont {Nazarenko}},\ and\ \bibinfo {author} {\bibfnamefont {S.}~\bibnamefont {Residori}},\ }\href {https://doi.org/10.1364/JOSAB.26.002280} {\bibfield  {journal} {\bibinfo  {journal} {Journal of the Optical Society of America B}\ }\textbf {\bibinfo {volume} {26}},\ \bibinfo {pages} {2280} (\bibinfo {year} {2009})}\BibitemShut {NoStop}%
\bibitem [{\citenamefont {Laurie}\ \emph {et~al.}(2012)\citenamefont {Laurie}, \citenamefont {Bortolozzo}, \citenamefont {Nazarenko},\ and\ \citenamefont {Residori}}]{laurie_one-dimensional_2012}%
  \BibitemOpen
  \bibfield  {author} {\bibinfo {author} {\bibfnamefont {J.}~\bibnamefont {Laurie}}, \bibinfo {author} {\bibfnamefont {U.}~\bibnamefont {Bortolozzo}}, \bibinfo {author} {\bibfnamefont {S.}~\bibnamefont {Nazarenko}},\ and\ \bibinfo {author} {\bibfnamefont {S.}~\bibnamefont {Residori}},\ }\href {https://doi.org/10.1016/j.physrep.2012.01.004} {\bibfield  {journal} {\bibinfo  {journal} {Physics Reports}\ }\bibinfo {series} {One-{Dimensional} {Optical} {Wave} {Turbulence}: {Experiment} and {Theory}},\ \textbf {\bibinfo {volume} {514}},\ \bibinfo {pages} {121} (\bibinfo {year} {2012})}\BibitemShut {NoStop}%
\bibitem [{\citenamefont {Coll{\'e}aux}\ \emph {et~al.}(2024)\citenamefont {Coll{\'e}aux}, \citenamefont {Skipp}, \citenamefont {Nazarenko},\ and\ \citenamefont {Laurie}}]{Colleaux2024boundstate}%
  \BibitemOpen
  \bibfield  {author} {\bibinfo {author} {\bibfnamefont {C.}~\bibnamefont {Coll{\'e}aux}}, \bibinfo {author} {\bibfnamefont {J.}~\bibnamefont {Skipp}}, \bibinfo {author} {\bibfnamefont {S.}~\bibnamefont {Nazarenko}},\ and\ \bibinfo {author} {\bibfnamefont {J.}~\bibnamefont {Laurie}},\ }\href {https://doi.org/https://doi.org/10.48550/arXiv.2410.12507} {\bibinfo {title} {A bound state attractor in optical turbulence}} (\bibinfo {year} {2024})\BibitemShut {NoStop}%
\bibitem [{\citenamefont {Agafontsev}\ \emph {et~al.}(2023)\citenamefont {Agafontsev}, \citenamefont {Gelash}, \citenamefont {Mullyadzhanov},\ and\ \citenamefont {Zakharov}}]{agafontsev_bound-state_2023}%
  \BibitemOpen
  \bibfield  {author} {\bibinfo {author} {\bibfnamefont {D.~S.}\ \bibnamefont {Agafontsev}}, \bibinfo {author} {\bibfnamefont {A.~A.}\ \bibnamefont {Gelash}}, \bibinfo {author} {\bibfnamefont {R.~I.}\ \bibnamefont {Mullyadzhanov}},\ and\ \bibinfo {author} {\bibfnamefont {V.~E.}\ \bibnamefont {Zakharov}},\ }\href {https://doi.org/10.1016/j.chaos.2022.112951} {\bibfield  {journal} {\bibinfo  {journal} {Chaos, Solitons \& Fractals}\ }\textbf {\bibinfo {volume} {166}},\ \bibinfo {pages} {112951} (\bibinfo {year} {2023})}\BibitemShut {NoStop}%
\bibitem [{\citenamefont {Suret}\ \emph {et~al.}(2024)\citenamefont {Suret}, \citenamefont {Randoux}, \citenamefont {Gelash}, \citenamefont {Agafontsev}, \citenamefont {Doyon},\ and\ \citenamefont {El}}]{suret_soliton_2024}%
  \BibitemOpen
  \bibfield  {author} {\bibinfo {author} {\bibfnamefont {P.}~\bibnamefont {Suret}}, \bibinfo {author} {\bibfnamefont {S.}~\bibnamefont {Randoux}}, \bibinfo {author} {\bibfnamefont {A.}~\bibnamefont {Gelash}}, \bibinfo {author} {\bibfnamefont {D.}~\bibnamefont {Agafontsev}}, \bibinfo {author} {\bibfnamefont {B.}~\bibnamefont {Doyon}},\ and\ \bibinfo {author} {\bibfnamefont {G.}~\bibnamefont {El}},\ }\href {https://doi.org/10.1103/PhysRevE.109.061001} {\bibfield  {journal} {\bibinfo  {journal} {Physical Review E}\ }\textbf {\bibinfo {volume} {109}},\ \bibinfo {pages} {061001} (\bibinfo {year} {2024})},\ \bibinfo {note} {publisher: American Physical Society}\BibitemShut {NoStop}%
\bibitem [{\citenamefont {Soto-Crespo}\ \emph {et~al.}(2016)\citenamefont {Soto-Crespo}, \citenamefont {Devine},\ and\ \citenamefont {Akhmediev}}]{soto-crespo_integrable_2016}%
  \BibitemOpen
  \bibfield  {author} {\bibinfo {author} {\bibfnamefont {J.~M.}\ \bibnamefont {Soto-Crespo}}, \bibinfo {author} {\bibfnamefont {N.}~\bibnamefont {Devine}},\ and\ \bibinfo {author} {\bibfnamefont {N.}~\bibnamefont {Akhmediev}},\ }\href {https://doi.org/10.1103/PhysRevLett.116.103901} {\bibfield  {journal} {\bibinfo  {journal} {Physical Review Letters}\ }\textbf {\bibinfo {volume} {116}},\ \bibinfo {pages} {103901} (\bibinfo {year} {2016})}\BibitemShut {NoStop}%
\bibitem [{\citenamefont {Randoux}\ \emph {et~al.}(2016)\citenamefont {Randoux}, \citenamefont {Walczak}, \citenamefont {Onorato},\ and\ \citenamefont {Suret}}]{randoux_nonlinear_2016}%
  \BibitemOpen
  \bibfield  {author} {\bibinfo {author} {\bibfnamefont {S.}~\bibnamefont {Randoux}}, \bibinfo {author} {\bibfnamefont {P.}~\bibnamefont {Walczak}}, \bibinfo {author} {\bibfnamefont {M.}~\bibnamefont {Onorato}},\ and\ \bibinfo {author} {\bibfnamefont {P.}~\bibnamefont {Suret}},\ }\href {https://doi.org/10.1016/j.physd.2016.04.001} {\bibfield  {journal} {\bibinfo  {journal} {Physica D: Nonlinear Phenomena}\ }\bibinfo {series} {Dispersive {Hydrodynamics}},\ \textbf {\bibinfo {volume} {333}},\ \bibinfo {pages} {323} (\bibinfo {year} {2016})}\BibitemShut {NoStop}%
\bibitem [{\citenamefont {Gelash}\ and\ \citenamefont {Agafontsev}(2018)}]{gelash_strongly_2018}%
  \BibitemOpen
  \bibfield  {author} {\bibinfo {author} {\bibfnamefont {A.~A.}\ \bibnamefont {Gelash}}\ and\ \bibinfo {author} {\bibfnamefont {D.~S.}\ \bibnamefont {Agafontsev}},\ }\href {https://doi.org/10.1103/PhysRevE.98.042210} {\bibfield  {journal} {\bibinfo  {journal} {Physical Review E}\ }\textbf {\bibinfo {volume} {98}},\ \bibinfo {pages} {042210} (\bibinfo {year} {2018})},\ \bibinfo {note} {publisher: American Physical Society}\BibitemShut {NoStop}%
\bibitem [{\citenamefont {Randoux}\ \emph {et~al.}(2014)\citenamefont {Randoux}, \citenamefont {Walczak}, \citenamefont {Onorato},\ and\ \citenamefont {Suret}}]{randoux_intermittency_2014}%
  \BibitemOpen
  \bibfield  {author} {\bibinfo {author} {\bibfnamefont {S.}~\bibnamefont {Randoux}}, \bibinfo {author} {\bibfnamefont {P.}~\bibnamefont {Walczak}}, \bibinfo {author} {\bibfnamefont {M.}~\bibnamefont {Onorato}},\ and\ \bibinfo {author} {\bibfnamefont {P.}~\bibnamefont {Suret}},\ }\href {https://doi.org/10.1103/PhysRevLett.113.113902} {\bibfield  {journal} {\bibinfo  {journal} {Physical Review Letters}\ }\textbf {\bibinfo {volume} {113}},\ \bibinfo {pages} {113902} (\bibinfo {year} {2014})}\BibitemShut {NoStop}%
\bibitem [{\citenamefont {Roberti}\ \emph {et~al.}(2019)\citenamefont {Roberti}, \citenamefont {El}, \citenamefont {Randoux},\ and\ \citenamefont {Suret}}]{roberti_early_2019}%
  \BibitemOpen
  \bibfield  {author} {\bibinfo {author} {\bibfnamefont {G.}~\bibnamefont {Roberti}}, \bibinfo {author} {\bibfnamefont {G.}~\bibnamefont {El}}, \bibinfo {author} {\bibfnamefont {S.}~\bibnamefont {Randoux}},\ and\ \bibinfo {author} {\bibfnamefont {P.}~\bibnamefont {Suret}},\ }\href {https://doi.org/10.1103/PhysRevE.100.032212} {\bibfield  {journal} {\bibinfo  {journal} {Physical Review E}\ }\textbf {\bibinfo {volume} {100}},\ \bibinfo {pages} {032212} (\bibinfo {year} {2019})},\ \bibinfo {note} {publisher: American Physical Society}\BibitemShut {NoStop}%
\bibitem [{\citenamefont {Congy}\ \emph {et~al.}(2024)\citenamefont {Congy}, \citenamefont {El}, \citenamefont {Roberti}, \citenamefont {Tovbis}, \citenamefont {Randoux},\ and\ \citenamefont {Suret}}]{congy_statistics_2024}%
  \BibitemOpen
  \bibfield  {author} {\bibinfo {author} {\bibfnamefont {T.}~\bibnamefont {Congy}}, \bibinfo {author} {\bibfnamefont {G.~A.}\ \bibnamefont {El}}, \bibinfo {author} {\bibfnamefont {G.}~\bibnamefont {Roberti}}, \bibinfo {author} {\bibfnamefont {A.}~\bibnamefont {Tovbis}}, \bibinfo {author} {\bibfnamefont {S.}~\bibnamefont {Randoux}},\ and\ \bibinfo {author} {\bibfnamefont {P.}~\bibnamefont {Suret}},\ }\href {https://doi.org/10.1103/PhysRevLett.132.207201} {\bibfield  {journal} {\bibinfo  {journal} {Physical Review Letters}\ }\textbf {\bibinfo {volume} {132}},\ \bibinfo {pages} {207201} (\bibinfo {year} {2024})},\ \bibinfo {note} {publisher: American Physical Society}\BibitemShut {NoStop}%
\bibitem [{\citenamefont {Castillo}\ \emph {et~al.}(1996)\citenamefont {Castillo}, \citenamefont {S{\'a}nchez-Mondrag{\'o}n},\ and\ \citenamefont {Stepanov}}]{castillo_formation_1996}%
  \BibitemOpen
  \bibfield  {author} {\bibinfo {author} {\bibfnamefont {M.~D.~I.}\ \bibnamefont {Castillo}}, \bibinfo {author} {\bibfnamefont {J.~J.}\ \bibnamefont {S{\'a}nchez-Mondrag{\'o}n}},\ and\ \bibinfo {author} {\bibfnamefont {S.}~\bibnamefont {Stepanov}},\ }\href {https://doi.org/10.1364/OL.21.001622} {\bibfield  {journal} {\bibinfo  {journal} {Optics Letters}\ }\textbf {\bibinfo {volume} {21}},\ \bibinfo {pages} {1622} (\bibinfo {year} {1996})},\ \bibinfo {note} {publisher: Optica Publishing Group}\BibitemShut {NoStop}%
\bibitem [{\citenamefont {Bekenstein}\ \emph {et~al.}(2015)\citenamefont {Bekenstein}, \citenamefont {Schley}, \citenamefont {Mutzafi}, \citenamefont {Rotschild},\ and\ \citenamefont {Segev}}]{bekenstein_optical_2015}%
  \BibitemOpen
  \bibfield  {author} {\bibinfo {author} {\bibfnamefont {R.}~\bibnamefont {Bekenstein}}, \bibinfo {author} {\bibfnamefont {R.}~\bibnamefont {Schley}}, \bibinfo {author} {\bibfnamefont {M.}~\bibnamefont {Mutzafi}}, \bibinfo {author} {\bibfnamefont {C.}~\bibnamefont {Rotschild}},\ and\ \bibinfo {author} {\bibfnamefont {M.}~\bibnamefont {Segev}},\ }\href {https://doi.org/10.1038/nphys3451} {\bibfield  {journal} {\bibinfo  {journal} {Nature Physics}\ }\textbf {\bibinfo {volume} {11}},\ \bibinfo {pages} {872} (\bibinfo {year} {2015})}\BibitemShut {NoStop}%
\bibitem [{\citenamefont {Roger}\ \emph {et~al.}(2016)\citenamefont {Roger}, \citenamefont {Maitland}, \citenamefont {Wilson}, \citenamefont {Westerberg}, \citenamefont {Vocke}, \citenamefont {Wright},\ and\ \citenamefont {Faccio}}]{roger_optical_2016}%
  \BibitemOpen
  \bibfield  {author} {\bibinfo {author} {\bibfnamefont {T.}~\bibnamefont {Roger}}, \bibinfo {author} {\bibfnamefont {C.}~\bibnamefont {Maitland}}, \bibinfo {author} {\bibfnamefont {K.}~\bibnamefont {Wilson}}, \bibinfo {author} {\bibfnamefont {N.}~\bibnamefont {Westerberg}}, \bibinfo {author} {\bibfnamefont {D.}~\bibnamefont {Vocke}}, \bibinfo {author} {\bibfnamefont {E.~M.}\ \bibnamefont {Wright}},\ and\ \bibinfo {author} {\bibfnamefont {D.}~\bibnamefont {Faccio}},\ }\href {https://doi.org/10.1038/ncomms13492} {\bibfield  {journal} {\bibinfo  {journal} {Nature Communications}\ }\textbf {\bibinfo {volume} {7}},\ \bibinfo {pages} {13492} (\bibinfo {year} {2016})}\BibitemShut {NoStop}%
\bibitem [{\citenamefont {Conti}\ \emph {et~al.}(2003)\citenamefont {Conti}, \citenamefont {Peccianti},\ and\ \citenamefont {Assanto}}]{conti_route_2003}%
  \BibitemOpen
  \bibfield  {author} {\bibinfo {author} {\bibfnamefont {C.}~\bibnamefont {Conti}}, \bibinfo {author} {\bibfnamefont {M.}~\bibnamefont {Peccianti}},\ and\ \bibinfo {author} {\bibfnamefont {G.}~\bibnamefont {Assanto}},\ }\href {https://doi.org/10.1103/PhysRevLett.91.073901} {\bibfield  {journal} {\bibinfo  {journal} {Physical Review Letters}\ }\textbf {\bibinfo {volume} {91}},\ \bibinfo {pages} {073901} (\bibinfo {year} {2003})}\BibitemShut {NoStop}%
\bibitem [{\citenamefont {Conti}\ \emph {et~al.}(2004)\citenamefont {Conti}, \citenamefont {Peccianti},\ and\ \citenamefont {Assanto}}]{conti_observation_2004}%
  \BibitemOpen
  \bibfield  {author} {\bibinfo {author} {\bibfnamefont {C.}~\bibnamefont {Conti}}, \bibinfo {author} {\bibfnamefont {M.}~\bibnamefont {Peccianti}},\ and\ \bibinfo {author} {\bibfnamefont {G.}~\bibnamefont {Assanto}},\ }\href {https://doi.org/10.1103/PhysRevLett.92.113902} {\bibfield  {journal} {\bibinfo  {journal} {Physical Review Letters}\ }\textbf {\bibinfo {volume} {92}},\ \bibinfo {pages} {113902} (\bibinfo {year} {2004})}\BibitemShut {NoStop}%
\bibitem [{\citenamefont {Zakharov}\ \emph {et~al.}(1992)\citenamefont {Zakharov}, \citenamefont {L{\textquoteright}vov},\ and\ \citenamefont {Falkovich}}]{zakharov_kolmogorov_1992}%
  \BibitemOpen
  \bibfield  {author} {\bibinfo {author} {\bibfnamefont {V.~E.}\ \bibnamefont {Zakharov}}, \bibinfo {author} {\bibfnamefont {V.~S.}\ \bibnamefont {L{\textquoteright}vov}},\ and\ \bibinfo {author} {\bibfnamefont {G.}~\bibnamefont {Falkovich}},\ }\href {http://www.springer.com/physics/complexity/book/978-3-642-50054-1} {\emph {\bibinfo {title} {Kolmogorov {Spectra} of {Turbulence} {I} - {Wave} {Turbulence}}}},\ Springer {Series} in {Nonlinear} {Dynamics}\ (\bibinfo  {publisher} {Springer Berlin Heidelberg},\ \bibinfo {year} {1992})\BibitemShut {NoStop}%
\bibitem [{\citenamefont {Nazarenko}(2011)}]{nazarenko_wave_2011}%
  \BibitemOpen
  \bibfield  {author} {\bibinfo {author} {\bibfnamefont {S.}~\bibnamefont {Nazarenko}},\ }\href {http://www.springer.com/physics/statistical+physics+%26+dynamical+systems/book/978-3-642-15941-1} {\emph {\bibinfo {title} {Wave {Turbulence}}}},\ \bibinfo {series} {Lecture {Notes} in {Physics}}\ No.\ \bibinfo {number} {825}\ (\bibinfo  {publisher} {Springer Berlin Heidelberg},\ \bibinfo {year} {2011})\BibitemShut {NoStop}%
\bibitem [{\citenamefont {Krasitskii}(1994)}]{krasitskii_reduced_1994}%
  \BibitemOpen
  \bibfield  {author} {\bibinfo {author} {\bibfnamefont {V.~P.}\ \bibnamefont {Krasitskii}},\ }\href {https://doi.org/10.1017/S0022112094004350} {\bibfield  {journal} {\bibinfo  {journal} {Journal of Fluid Mechanics}\ }\textbf {\bibinfo {volume} {272}},\ \bibinfo {pages} {1} (\bibinfo {year} {1994})}\BibitemShut {NoStop}%
\bibitem [{\citenamefont {Dyachenko}\ \emph {et~al.}(1995)\citenamefont {Dyachenko}, \citenamefont {Lvov},\ and\ \citenamefont {Zakharov}}]{dyachenko_five-wave_1995}%
  \BibitemOpen
  \bibfield  {author} {\bibinfo {author} {\bibfnamefont {A.}~\bibnamefont {Dyachenko}}, \bibinfo {author} {\bibfnamefont {Y.}~\bibnamefont {Lvov}},\ and\ \bibinfo {author} {\bibfnamefont {V.}~\bibnamefont {Zakharov}},\ }\href {https://doi.org/10.1016/0167-2789(95)00168-4} {\bibfield  {journal} {\bibinfo  {journal} {Physica D: Nonlinear Phenomena}\ }\textbf {\bibinfo {volume} {87}},\ \bibinfo {pages} {233} (\bibinfo {year} {1995})}\BibitemShut {NoStop}%
\bibitem [{\citenamefont {Kozik}\ and\ \citenamefont {Svistunov}(2004)}]{kozik_kelvin-wave_2004}%
  \BibitemOpen
  \bibfield  {author} {\bibinfo {author} {\bibfnamefont {E.}~\bibnamefont {Kozik}}\ and\ \bibinfo {author} {\bibfnamefont {B.}~\bibnamefont {Svistunov}},\ }\href {https://doi.org/10.1103/PhysRevLett.92.035301} {\bibfield  {journal} {\bibinfo  {journal} {Physical Review Letters}\ }\textbf {\bibinfo {volume} {92}},\ \bibinfo {pages} {035301} (\bibinfo {year} {2004})}\BibitemShut {NoStop}%
\bibitem [{\citenamefont {Laurie}\ \emph {et~al.}(2010)\citenamefont {Laurie}, \citenamefont {L{\textquoteright}vov}, \citenamefont {Nazarenko},\ and\ \citenamefont {Rudenko}}]{laurie_interaction_2010}%
  \BibitemOpen
  \bibfield  {author} {\bibinfo {author} {\bibfnamefont {J.}~\bibnamefont {Laurie}}, \bibinfo {author} {\bibfnamefont {V.~S.}\ \bibnamefont {L{\textquoteright}vov}}, \bibinfo {author} {\bibfnamefont {S.}~\bibnamefont {Nazarenko}},\ and\ \bibinfo {author} {\bibfnamefont {O.}~\bibnamefont {Rudenko}},\ }\href {https://doi.org/10.1103/PhysRevB.81.104526} {\bibfield  {journal} {\bibinfo  {journal} {Physical Review B}\ }\textbf {\bibinfo {volume} {81}},\ \bibinfo {pages} {104526} (\bibinfo {year} {2010})}\BibitemShut {NoStop}%
\bibitem [{\citenamefont {Fj{\o}rtoft}(1953)}]{fjortoft_changes_1953}%
  \BibitemOpen
  \bibfield  {author} {\bibinfo {author} {\bibfnamefont {R.}~\bibnamefont {Fj{\o}rtoft}},\ }\href {https://doi.org/10.1111/j.2153-3490.1953.tb01051.x} {\bibfield  {journal} {\bibinfo  {journal} {Tellus}\ }\textbf {\bibinfo {volume} {5}},\ \bibinfo {pages} {225} (\bibinfo {year} {1953})}\BibitemShut {NoStop}%
\bibitem [{\citenamefont {Skipp}\ \emph {et~al.}(2020)\citenamefont {Skipp}, \citenamefont {L'vov},\ and\ \citenamefont {Nazarenko}}]{skipp_wave_2020}%
  \BibitemOpen
  \bibfield  {author} {\bibinfo {author} {\bibfnamefont {J.}~\bibnamefont {Skipp}}, \bibinfo {author} {\bibfnamefont {V.}~\bibnamefont {L'vov}},\ and\ \bibinfo {author} {\bibfnamefont {S.}~\bibnamefont {Nazarenko}},\ }\href {https://doi.org/10.1103/PhysRevA.102.043318} {\bibfield  {journal} {\bibinfo  {journal} {Physical Review A}\ }\textbf {\bibinfo {volume} {102}},\ \bibinfo {pages} {043318} (\bibinfo {year} {2020})},\ \bibinfo {note} {publisher: American Physical Society}\BibitemShut {NoStop}%
\bibitem [{\citenamefont {Skipp}\ \emph {et~al.}(2023)\citenamefont {Skipp}, \citenamefont {Laurie},\ and\ \citenamefont {Nazarenko}}]{skipp_effective_2023}%
  \BibitemOpen
  \bibfield  {author} {\bibinfo {author} {\bibfnamefont {J.}~\bibnamefont {Skipp}}, \bibinfo {author} {\bibfnamefont {J.}~\bibnamefont {Laurie}},\ and\ \bibinfo {author} {\bibfnamefont {S.~V.}\ \bibnamefont {Nazarenko}},\ }\href {https://doi.org/10.1098/rspa.2023.0162} {\bibfield  {journal} {\bibinfo  {journal} {Proceedings of the Royal Society A: Mathematical, Physical and Engineering Sciences}\ }\textbf {\bibinfo {volume} {479}},\ \bibinfo {pages} {20230162} (\bibinfo {year} {2023})},\ \bibinfo {note} {publisher: Royal Society}\BibitemShut {NoStop}%
\bibitem [{\citenamefont {Horvai}\ \emph {et~al.}(2008)\citenamefont {Horvai}, \citenamefont {Nazarenko},\ and\ \citenamefont {Stein}}]{horvai_coalescence_2008}%
  \BibitemOpen
  \bibfield  {author} {\bibinfo {author} {\bibfnamefont {P.}~\bibnamefont {Horvai}}, \bibinfo {author} {\bibfnamefont {S.~V.}\ \bibnamefont {Nazarenko}},\ and\ \bibinfo {author} {\bibfnamefont {T.~H.~M.}\ \bibnamefont {Stein}},\ }\href {https://doi.org/10.1007/s10955-007-9466-y} {\bibfield  {journal} {\bibinfo  {journal} {Journal of Statistical Physics}\ }\textbf {\bibinfo {volume} {130}},\ \bibinfo {pages} {1177} (\bibinfo {year} {2008})}\BibitemShut {NoStop}%
\bibitem [{\citenamefont {Pearson}\ \emph {et~al.}(1984)\citenamefont {Pearson}, \citenamefont {Valioulis},\ and\ \citenamefont {List}}]{pearson1984monte}%
  \BibitemOpen
  \bibfield  {author} {\bibinfo {author} {\bibfnamefont {H.}~\bibnamefont {Pearson}}, \bibinfo {author} {\bibfnamefont {I.}~\bibnamefont {Valioulis}},\ and\ \bibinfo {author} {\bibfnamefont {E.}~\bibnamefont {List}},\ }\href@noop {} {\bibfield  {journal} {\bibinfo  {journal} {Journal of Fluid Mechanics}\ }\textbf {\bibinfo {volume} {143}},\ \bibinfo {pages} {367} (\bibinfo {year} {1984})}\BibitemShut {NoStop}%
\bibitem [{\citenamefont {Boffetta}\ and\ \citenamefont {Ecke}(2012)}]{boffetta_two-dimensional_2012}%
  \BibitemOpen
  \bibfield  {author} {\bibinfo {author} {\bibfnamefont {G.}~\bibnamefont {Boffetta}}\ and\ \bibinfo {author} {\bibfnamefont {R.~E.}\ \bibnamefont {Ecke}},\ }\href {https://doi.org/10.1146/annurev-fluid-120710-101240} {\bibfield  {journal} {\bibinfo  {journal} {Annual Review of Fluid Mechanics}\ }\textbf {\bibinfo {volume} {44}},\ \bibinfo {pages} {427} (\bibinfo {year} {2012})}\BibitemShut {NoStop}%
\bibitem [{\citenamefont {Picozzi}\ and\ \citenamefont {Garnier}(2011)}]{picozzi_incoherent_2011}%
  \BibitemOpen
  \bibfield  {author} {\bibinfo {author} {\bibfnamefont {A.}~\bibnamefont {Picozzi}}\ and\ \bibinfo {author} {\bibfnamefont {J.}~\bibnamefont {Garnier}},\ }\href {https://doi.org/10.1103/PhysRevLett.107.233901} {\bibfield  {journal} {\bibinfo  {journal} {Physical Review Letters}\ }\textbf {\bibinfo {volume} {107}},\ \bibinfo {pages} {233901} (\bibinfo {year} {2011})}\BibitemShut {NoStop}%
\bibitem [{\citenamefont {Garnier}\ \emph {et~al.}(2021)\citenamefont {Garnier}, \citenamefont {Baudin}, \citenamefont {Fusaro},\ and\ \citenamefont {Picozzi}}]{garnier_incoherent_2021}%
  \BibitemOpen
  \bibfield  {author} {\bibinfo {author} {\bibfnamefont {J.}~\bibnamefont {Garnier}}, \bibinfo {author} {\bibfnamefont {K.}~\bibnamefont {Baudin}}, \bibinfo {author} {\bibfnamefont {A.}~\bibnamefont {Fusaro}},\ and\ \bibinfo {author} {\bibfnamefont {A.}~\bibnamefont {Picozzi}},\ }\href {https://doi.org/10.1103/PhysRevE.104.054205} {\bibfield  {journal} {\bibinfo  {journal} {Physical Review E}\ }\textbf {\bibinfo {volume} {104}},\ \bibinfo {pages} {054205} (\bibinfo {year} {2021})},\ \bibinfo {note} {publisher: American Physical Society}\BibitemShut {NoStop}%
\bibitem [{\citenamefont {Boffetta}\ \emph {et~al.}(2009)\citenamefont {Boffetta}, \citenamefont {Celani}, \citenamefont {Dezzani}, \citenamefont {Laurie},\ and\ \citenamefont {Nazarenko}}]{boffetta_modeling_2009}%
  \BibitemOpen
  \bibfield  {author} {\bibinfo {author} {\bibfnamefont {G.}~\bibnamefont {Boffetta}}, \bibinfo {author} {\bibfnamefont {A.}~\bibnamefont {Celani}}, \bibinfo {author} {\bibfnamefont {D.}~\bibnamefont {Dezzani}}, \bibinfo {author} {\bibfnamefont {J.}~\bibnamefont {Laurie}},\ and\ \bibinfo {author} {\bibfnamefont {S.}~\bibnamefont {Nazarenko}},\ }\href {https://doi.org/10.1007/s10909-009-9895-x} {\bibfield  {journal} {\bibinfo  {journal} {Journal of Low Temperature Physics}\ }\textbf {\bibinfo {volume} {156}},\ \bibinfo {pages} {193} (\bibinfo {year} {2009})}\BibitemShut {NoStop}%
\bibitem [{\citenamefont {M{\"u}ller}\ and\ \citenamefont {Krstulovic}(2020)}]{muller_kolmogorov_2020}%
  \BibitemOpen
  \bibfield  {author} {\bibinfo {author} {\bibfnamefont {N.~P.}\ \bibnamefont {M{\"u}ller}}\ and\ \bibinfo {author} {\bibfnamefont {G.}~\bibnamefont {Krstulovic}},\ }\href {https://doi.org/10.1103/PhysRevB.102.134513} {\bibfield  {journal} {\bibinfo  {journal} {Physical Review B}\ }\textbf {\bibinfo {volume} {102}},\ \bibinfo {pages} {134513} (\bibinfo {year} {2020})},\ \bibinfo {note} {publisher: American Physical Society}\BibitemShut {NoStop}%
\bibitem [{\citenamefont {Dyachenko}\ \emph {et~al.}(1992)\citenamefont {Dyachenko}, \citenamefont {Newell}, \citenamefont {Pushkarev},\ and\ \citenamefont {Zakharov}}]{dyachenko_optical_1992}%
  \BibitemOpen
  \bibfield  {author} {\bibinfo {author} {\bibfnamefont {S.}~\bibnamefont {Dyachenko}}, \bibinfo {author} {\bibfnamefont {A.~C.}\ \bibnamefont {Newell}}, \bibinfo {author} {\bibfnamefont {A.}~\bibnamefont {Pushkarev}},\ and\ \bibinfo {author} {\bibfnamefont {V.~E.}\ \bibnamefont {Zakharov}},\ }\href {https://doi.org/10.1016/0167-2789(92)90090-A} {\bibfield  {journal} {\bibinfo  {journal} {Physica D: Nonlinear Phenomena}\ }\textbf {\bibinfo {volume} {57}},\ \bibinfo {pages} {96} (\bibinfo {year} {1992})}\BibitemShut {NoStop}%
\end{thebibliography}%

\onecolumngrid
\pagebreak

\begin{center}
\textbf{\large Supplemental Material: Semi-local one-dimensional optical wave turbulence}
\end{center}
\setcounter{equation}{0}
\setcounter{figure}{0}
\setcounter{table}{0}
\setcounter{page}{1}
\makeatletter
\renewcommand{\theequation}{S\arabic{equation}}
\renewcommand{\thefigure}{S\arabic{figure}}
\renewcommand{\thetable}{S\arabic{table}}

\section{Canonical transformation}

\noindent In this section, we provide details of the canonical transformation required to remove the leading four-wave non-resonant interactions arising from the cubic nonlinearity of the 1D SHE. We find that this leads to the resonant interaction becoming of six-wave type in the leading order. The 1D SHE belongs to a class of 1D models in which the dispersion relation $\omega_{\bf k}\propto {k}^\beta$, with $\beta>1$, prevents non-trivial solutions of the four-wave resonant condition that defines $2\leftrightarrow 2$ wave mixing. Subsequently, one must apply a canonical transformation to the wave amplitude variable $\hat{\psi}_{\bf k}$ to recast Eq.~\eqref{eq:she_fourier} into a form where the leading nonlinear is resonant. The procedure is well-documented for this class of 1D systems arising in nonlinear optics~\cite{bortolozzo_optical_2009,laurie_one-dimensional_2012} and Kelvin waves in superfluid helium-4~\cite{laurie_interaction_2010}, where the general result is the same as for the 1D SHE. Briefly, the approach is to define a one-parametric quasi-linear transformation of the form of a Taylor series in a parameter $\tau$,
\begin{align}\label{eq:ct}
\hat{\psi}_{\bf k}(t)=c_{\bf k}(t,\tau)\bigg|_{\tau=0}+ \tau \left.\frac{\partial c_{\bf k}(t,\tau)}{\partial \tau}\right|_{\tau=0}+ \frac{\tau^2}{2} \left.\frac{\partial^2 c_{\bf k}(t,\tau)}{\partial \tau^2}\right|_{\tau=0}+ \cdots.
\end{align}
The parameter $\tau$ is treated as an ``auxiliary time'' so that the variable $c_{\bf k}(t,\tau)$ satisfies Hamiltonian's equation with an auxiliary Hamiltonian $H_{\rm aux}$ consisting of only interacting wave terms of three-wave order and above, but no quadratic part. This procedure ensures that the transformation remains canonical, with the canonical expansion coefficients computed via $\partial c_{\bf k}/\partial \tau = -
i\partial H_{\rm aux}/\partial c_{\bf k}^*$, $\partial^2 c_{\bf k}/\partial \tau^2 = -i \partial/\partial \tau\left[\partial H_{\rm aux}/\partial c_{\bf k}^*\right]$, etc. Substitution of~\eqref{eq:ct} into the 1D SHE~\eqref{eq:she_fourier}, with the requirement that we eliminate the cubic terms that describe non-resonant four-wave mixing, defines the (four-wave and six-wave) coefficients of the auxiliary Hamiltonian $H_{\rm aux}$ and leads to an additional contribution arising at the six-wave term. Ultimately, this procedure transforms Eq.~\eqref{eq:she_fourier} to an equation of motion for $c_{\bf k}(t)$:
\begin{align}\label{eq:c_dot}
i\frac{\partial c_{\bf k}}{\partial t} =\omega_{\bf k}c_{\bf k} + \sum_{{\bf k}_1, {\bf k}_2, {\bf k}_3, {\bf k}_4, {\bf k}_5} W^{1,2,3}_{4,5,{\bf k}}c_{1}c_{2}c_{3}c_{4}^*c_{5}^* \delta^{1,2,3}_{4,5,{\bf k}},
\end{align}
with the same linear-wave frequency $\omega_{\bf k} = k^2/2$ and a six-wave interaction coefficient $W^{1,2,3}_{4,5,6}$ defined in terms of the original $T^{1,2}_{3,4}$ as given in~\eqref{eq:T}:
\begin{align}
W^{1,2,3}_{4,5,6} =\frac{1}{24} \sum^3_{\substack{i,j,m =1\\ i\neq j \neq m\neq i}} \sum^6_{\substack{p,q,r =4\\ p\neq q \neq r\neq p}} \left[ \frac{1}{\omega^{i,j}_{r,i+j-r}} -\frac{1}{\omega^{p+q-m,m}_{p.q}} \right]T^{i,j}_{r,i+j-r}T^{q+p-m,m}_{q,p},
\end{align}
where 
$\omega^{i,j}_{r,i+j-r} =
\omega_{{\bf k}_i} + \omega_{{\bf k}_j} -
\omega_{{\bf k}_r} - \omega_{{\bf k}_i + {\bf k}_j - {\bf k}_r}
$, etc. 
The essential point is that the nonlinear term in Eq.~\eqref{eq:c_dot} describes resonant six-wave mixing that arises due to a coupling of two non-resonant four-wave interactions. It is from this equation of motion that one applies the theory of wave turbulence to derive the kinetic equation~\eqref{eq:kinetic}.

\section{Derivation of the SLAM}

\noindent To derive the 1D SLAM, we begin by following a similar strategy to that for the 2D SLAM~\cite{skipp_effective_2023} by multiplying the WKE~\eqref{eq:kinetic} by an arbitrary test function $\varphi_{\bf k}=\varphi({\bf k})$ and integrating with respect to ${\bf k}$.
Using the symmetries of the collision integral, we arrive at
\begin{align}\label{eq:wke_test}
\int \varphi_{\bf k}\frac{\partial n_{\bf k}}{\partial t} {\rm d}{\bf k}= 4\pi\int & \left|W^{1,2,3}_{4,5,{\bf k}}\right|^2
\left[ \frac{1}{n_{{\bf k}}} +\frac{1}{n_5}+\frac{1}{n_4}-\frac{1}{n_1}-\frac{1}{n_2}-\frac{1}{n_3} \right]\  n_1n_2n_3n_4n_5n_{\bf k}\nonumber \\
&\times  \left[\varphi_{\bf k} + \varphi_5 + \varphi_4 - \varphi_3 - \varphi_2 - \varphi_1 \right] \delta^{1,2,3}_{4,5,{\bf k}}\ \delta(\omega^{1,2,3}_{4,5,{\bf k}})\ {\rm d}{\bf k}_1 {\rm d}{\bf k}_2 {\rm d}{\bf k}_3 {\rm d}{\bf k}_4 {\rm d}{\bf k}_5 {\rm d}{\bf k}.
\end{align}
We now take the semi-local limit. Recalling from the main text, this limit naturally arises if one considers $\Lambda \ll k^2$; then the interaction coefficient $W^{1,2,3}_{4,5,{\bf k}}$ becomes sharply peaked when three pairs of wavenumbers become close, i.e. ${\bf k}_1 \approx {\bf k}_4$, ${\bf k}_2 \approx {\bf k}_5$,  and ${\bf k}_3 \approx{\bf k}$. Here, we assume that the values of each pair, ${\bf k}_4$, ${\bf k}_5$ and ${\bf k}$ remain distinct. This is in contrast to the procedure of deriving differential approximation models~\cite{dyachenko_optical_1992}, where one assumes that all interactions are super-local. To formalise this derivation we consider the variables ${\bf p} = {\bf k}_1 - {\bf k}_4$, ${\bf q}={\bf k}_2-{\bf k}_5$, and consider the limit when $p \ll k_1,k_4$, $q \ll k_2,k_5$. Our goal is to reduce the dimensionality of the WKE and integrate out the divergence associated with the $\Lambda \to 0$ limit.

We first consider the Dirac delta function of wavenumbers, which leads to $\delta^{1,2,3}_{4,5,{\bf k}} = \delta \left({\bf p}+{\bf q}+ ({\bf k}_3-{\bf k})\right)$, which implies ${\bf k}_3 - {\bf k} = - {\bf p} - {\bf q}$. The Dirac delta function involving frequencies gives $\omega^{1,2,3}_{4,5,{\bf k}} = {\bf p}({\bf k}_1-{\bf k}) + {\bf q}({\bf k}_2-{\bf k}) + o(p) + o(q)$, thus at the leading order in $p$ and $q$ we have $\delta(\omega^{1,2,3}_{4,5,{\bf k}}) = \delta\left({\bf p}({\bf k}_1-{\bf k}) + {\bf q}({\bf k}_2 - {\bf k})\right)$.

The six-wave interaction coefficient $W^{1,2,3}_{4,5,{\bf k}}$ can then be simplified in a similar manner by expansion in the semi-local approximation limit, with the substitutions ${\bf k}_4={\bf k}_1-{\bf p}$, ${\bf k}_5={\bf k}_2-{\bf q}$, and ${\bf k}_3={\bf k}-{\bf p}-{\bf q}$. 
\begin{align*}
    {W}^{1,2,3}_{4,5,{\bf k}} = {W}^{1,2,{\bf k} - {\bf p}-{\bf q}}_{1-{\bf p},2-{\bf q},{\bf k}}&= \left(\frac{{\bf p}}{{\bf q}({\bf k}-{\bf k}_2)^2} +  \frac{{\bf q}}{{\bf p}({\bf k}-{\bf k}_1)^2} \right) \left( \frac{1}{{p}^2 + \Lambda}\right)\left(  \frac{1}{{q}^2 + \Lambda}\right) \\
    &- \left(  \frac{1}{({\bf k}_1-{\bf k}_2)^2} +  \frac{\bf q}{({\bf p}+{\bf q})({\bf k}-{\bf k}_1)^2} + \frac{{\bf p}}{{\bf q}({\bf k}_1-{\bf k}_2)^2}\right) \left(\frac{1}{{q}^2  + \Lambda}\right)  \left(\frac{1}{({\bf p}+{\bf q})^2 + \Lambda}\right) \\
    &-\left( \frac{1}{({\bf k}_1-{\bf k}_2)^2} +  \frac{{\bf p}}{({\bf p}+{\bf q})({\bf k}-{\bf k}_2)^2} + \frac{\bf q}{{\bf p}({\bf k}_1-{\bf k}_2)^2}\right) \left(\frac{1}{({\bf p}+{\bf q})^2 + \Lambda}\right) \left(\frac{1}{{p}^2  + \Lambda}\right).
\end{align*}
To derive the 1D SLAM, we substitute the above approximations into Eq.~\eqref{eq:wke_test}, and perform Taylor expansions in terms of $p$ and $q$ for the expressions involving the differences of $1/n_{{\bf k}_i}$ and test functions $\varphi_{{\bf k}_i}$. We then use the Dirac delta function of wavenumbers to integrate out ${\bf k}_2$, arriving at
\begin{align}
      \int \varphi_{\bf k} \frac{\partial {n}_{\bf k}}{\partial t} \, {\rm d} {\bf k}  
      &= 8\pi \int |W^{{\bf k}_1,{\bf k}_2,{\bf k}-{\bf p}-{\bf q}}_{{\bf k}_1-{\bf p},{\bf k}_2-{\bf q},{\bf k}}|^2  
       \left[ {\bf p} \frac{\partial}{\partial {\bf k}_1}\left(\frac{1}{n_{1}}\right) + {\bf q} \frac{\partial}{\partial {\bf k}_2}\left( \frac{1}{n_{2}}\right) + \left(\frac{1}{n_{{\bf k}-{\bf p}-{\bf q}}} -\frac{1}{n_{\bf k}}  \right)\right] \nonumber\\
       &\times
       n_{1} n_{2} n_{{\bf k}-{\bf p}-{\bf q}} n_{{\bf k}_1-{\bf p}} n_{{\bf k}_2-{\bf q}} n_{\bf k}
       \left[ {\bf p} \frac{\partial \varphi_1}{\partial {\bf k}_1} + {\bf q} \frac{\partial \varphi_2}
{\partial {\bf k}_2} + \left(\varphi_{{\bf k}-{\bf p}-{\bf q}}  -\varphi_{\bf k} \right) \right] \nonumber\\
& \times  \delta \left({\bf p}(2{\bf k}_1-{\bf p}) + {\bf q}(2{\bf k}_2+{\bf q}) + (-{\bf p}-{\bf q})(2{\bf k}-{\bf p}-{\bf q})\right)\, {\rm d} {\bf k}_1 {\rm d} {\bf k}_2 {\rm d} {\bf p} {\rm d} {\bf q} {\rm d} {\bf k}.
\end{align}

We perform a third Taylor expansion in the square brackets and the Dirac delta function, assuming $|{\bf p}+{\bf q}| \ll k$, retaining terms up to $o(p)$ and $o(q)$. We obtain
\begin{align}
      \int \varphi_{\bf k} \frac{\partial {n}_{\bf k}}{\partial t} \, {\rm d} {\bf k}  &= 4\pi \int |W^{{\bf k}_1,{\bf k}_2,{\bf k}-{\bf p}-{\bf q}}_{{\bf k}_1-{\bf p},{\bf k}_2-{\bf q},{\bf k}}|^2  
       \left[ {\bf p} \left(\frac{\partial}{\partial {\bf k}_1}\left(\frac{1}{n_{1}}\right)-\frac{\partial}{\partial {\bf k}}\left(\frac{1}{n_{\bf k}}\right)\right) + {\bf q} \left( \frac{\partial}{\partial {\bf k}_2}\left( \frac{1}{n_{2}}\right) - \frac{\partial}{\partial {\bf k}}\left(\frac{1}{n_{\bf k}}\right)\right) \right]\nonumber \\
       &\times n_{1} n_{2} n_{{\bf k}-{\bf p}-{\bf q}} n_{{\bf k}_1-{\bf p}} n_{{\bf k}_2-{\bf q}} n_{\bf k} \left[ {\bf p} \left(\frac{\partial \varphi_1}{\partial {\bf k}_1}- \frac{\partial \varphi_{\bf k}}{\partial {\bf k}}\right) + {\bf q} \left(\frac{\partial \varphi_2}
{\partial {\bf k}_2}  -\frac{\partial \varphi_{\bf k}}{\partial {\bf k}}\right)\right] \delta \left({\bf p}({\bf k}_1-{\bf k}) + {\bf q}({\bf k}_2-{\bf k})\right)\nonumber \\
&\times {\rm d} {\bf k}_1 {\rm d} {\bf k}_2 {\rm d} {\bf p} {\rm d} {\bf q} {\rm d} {\bf k}.
\end{align}
Now we integrate out variable ${\bf q}$ by using the final Dirac delta function with $\delta \left({\bf p}({\bf k}_1-{\bf k}) + {\bf q}({\bf k}_2-{\bf k})\right) = \delta \left({\bf q} + {\bf p}({\bf k}_1-{\bf k})/({\bf k}_2-{\bf k}) \right) / |{\bf k}_2 - {\bf k}|$, before finally Taylor expanding the wave action spectra $n_{i}$'s to the leading order in $p$:
\begin{align}\label{eq:slam_localisation}
      \int \varphi_{\bf k} \frac{\partial {n}_{\bf k}}{\partial t} \, {\rm d} {\bf k}  &= 4\pi \int \frac{p^2\left|W({\bf k}_1,{\bf k}_2,{\bf k},{\bf p})\right|^2}{|{\bf k}_2-{\bf k}|}
       \left[  \left(\frac{\partial}{\partial {\bf k}_1}\left(\frac{1}{n_{1}}\right)-\frac{\partial}{\partial {\bf k}}\left(\frac{1}{n_{\bf k}}\right)\right) + \frac{{\bf k}-{\bf k}_1}{{\bf k}_2-{\bf k}} \left( \frac{\partial}{\partial {\bf k}_2}\left( \frac{1}{n_{2}}\right) - \frac{\partial}{\partial {\bf k}}\left(\frac{1}{n_{\bf k}}\right)\right) \right]n_{1}^2 n_{2}^2 n_{{\bf k}}^2 \nonumber\\ 
       & \times  \left[ \left(\frac{\partial \varphi_1}{\partial {\bf k}_1}- \frac{\partial \varphi_{\bf k}}{\partial {\bf k}}\right) +  \frac{{\bf k}-{\bf k}_1}{{\bf k}_2-{\bf k}} \left(\frac{\partial \varphi_2}
{\partial {\bf k}_2}  -\frac{\partial \varphi_{\bf k}}{\partial {\bf k}}\right)\right] \, {\rm d} {\bf k}_1 {\rm d} {\bf k}_2 {\rm d} {\bf p} {\rm d} {\bf k},
\end{align}
where $W({\bf k}_1,{\bf k}_2,{\bf k},{\bf p})$ is the further simplified six-wave interaction coefficient involving the leading divergence in the semi-local limit:
\begin{align*}
W({\bf k}_1,{\bf k}_2,{\bf k},{\bf p}) &= -\left(  \frac{2}{({\bf k}-{\bf k}_2)({\bf k}-{\bf k}_1)} \right) \left( \frac{1}{{p}^2 + \Lambda}  \right)\left(\frac{1}{{p}^2 (\frac{{\bf k}-{\bf k}_1}{{\bf k}-{\bf k}_2})^2 + \Lambda} \right) \\
&-\left(  \frac{2}{({\bf k}_2-{\bf k}_1)({\bf k}_2-{\bf k})} \right) \left( \frac{1}{{p}^2 (\frac{{\bf k}_1-{\bf k}_2}{{\bf k}-{\bf k}_2})^2 + \Lambda}\right)  \left(\frac{1}{{p}^2  + \Lambda}\right)\\
&-\left(  \frac{2}{({\bf k}_1-{\bf k}_2)({\bf k}_1-{\bf k})} \right) \left( \frac{1}{{p}^2 (\frac{{\bf k}_1-{\bf k}_2}{{\bf k}-{\bf k}_2})^2 + \Lambda}  \right)\left(\frac{1}{{ p}^2 (\frac{{\bf k}-{\bf k}_1}{{\bf k}-{\bf k}_2})^2 + \Lambda}\right).
\end{align*}
It is now possible to directly integrate~\eqref{eq:slam_localisation} with respect to the variable ${\bf p}$ as we are able to write an explicit expression for terms only involving ${\bf p}$:
\begin{align}
\label{eq:Wtof}
\int p^2|W({\bf k}_1,{\bf k}_2,{\bf k},{\bf p})|^2 \ {\rm d} {\bf p}  = \frac{f({\bf k}_1,{\bf k}_2,{\bf k})}{\Lambda^{5/2}}, 
\end{align}
where the resulting integration gives
\begin{align}\label{eq:f}
f\left({\bf k}_1, {\bf k}_2, {\bf k}\right)
= -&\frac{ 
            18\pi({\bf k} - {\bf k}_2) 
        }
        {
            \left|{\bf k}-{\bf k}_2\right| ({\bf k}-{\bf k}_1)^3  ({\bf k}_1-{\bf k}_2)^3 
            ({\bf k}   + {\bf k}_1 - 2{\bf k}_2)^3  
            ({\bf k}   + {\bf k}_2 - 2{\bf k}_1)^3
            ({\bf k}_1 + {\bf k}_2 - 2{\bf k})^3  
        }
    \Biggl[ \nonumber \\
&-\biggl\{ 
    \left|{\bf k}-{\bf k}_1\right|  ({\bf k}-{\bf k}_1)^3  ({\bf k}+{\bf k}_1-2{\bf k}_2)^3 
       \Bigl(
            4{\bf k}^4  +  2{\bf k}^3 ({\bf k}_2 - 9{\bf k}_1)   
            + {\bf k}^2 (25{\bf k}_1^2 + 4{\bf k}_1{\bf k}_2 - 5{\bf k}_2^2) \nonumber\\
    &\quad+      {\bf k} (-18{\bf k}_1^3 + 4{\bf k}_1^2 {\bf k}_2 - 8{\bf k}_1{\bf k}_2^2 + 6{\bf k}_2^3)
            + 4{\bf k}_1^4 + 2{\bf k}_1^3 {\bf k}_2 - 5{\bf k}_1^2 {\bf k}_2^2 + 6{\bf k}_1 {\bf k}_2^3 - 3{\bf k}_2^4
        \Bigr)
\biggr\}\nonumber\\
&+\biggl\{ 
    \left|{\bf k}-{\bf k}_2\right|  ({\bf k}-{\bf k}_2)^3  ({\bf k}+{\bf k}_2-2{\bf k}_1)^3 
        \Bigl(
             4{\bf k}^4  +  2{\bf k}^3 ({\bf k}_1-9{\bf k}_2)
             + {\bf k}^2 (25{\bf k}_2^2 + 4{\bf k}_1{\bf k}_2 - 5{\bf k}_1^2) \nonumber \\
    &\quad+       {\bf k} (- 18{\bf k}_2^3 + 4{\bf k}_1{\bf k}_2^2 - 8{\bf k}_1^2{\bf k}_2 + 6{\bf k}_1^3)
             + 4{\bf k}_2^4 + 2{\bf k}_1{\bf k}_2^3 - 5{\bf k}_1^2{\bf k}_2^2  + 6{\bf k}_1^3{\bf k}_2 - 3{\bf k}_1^4 
        \Bigr)
\biggr\}\nonumber\\
&-\biggl\{ 
    \left|{\bf k}_1-{\bf k}_2\right|  ({\bf k}_1-{\bf k}_2)^3 ({\bf k}_1+{\bf k}_2-2{\bf k})^3 
        \Bigl(
           4{\bf k}_1^4  +  2{\bf k}_1^3 ({\bf k}-9{\bf k}_2)
             + {\bf k}_1^2 (25{\bf k}_2^2 + 4{\bf k}{\bf k}_2 - 5{\bf k}^2) \nonumber \\
    &\quad+       {\bf k}_1 (- 18{\bf k}_2^3 + 4{\bf k}{\bf k}_2^2 - 8{\bf k}^2{\bf k}_2 + 6{\bf k}^3)
             + 4{\bf k}_2^4 + 2{\bf k}{\bf k}_2^3 - 5{\bf k}^2{\bf k}_2^2 + 6{\bf k}^3{\bf k}_2 - 3{\bf k}^4 
        \Bigr)
\biggr\}
\Biggr].
\end{align}

Substituting Eqs.~\eqref{eq:Wtof} and~\eqref{eq:f} into Eq.~\eqref{eq:slam_localisation}, and using the symmetry of the resulting integral with respect to exchanging ${\bf k}_1, {\bf k}_2$ and ${\bf k}$, followed by integration by parts, and using the localisation theorem to remove the arbitrary test function $\varphi_{\bf k}$ from both sides, we obtain the 1D SLAM:

\begin{align}\label{eq:slam-derivation}
\frac{\partial n_{\bf k}}{\partial t} &=  \frac{1}{\Lambda^{5/2}}\frac{\partial}{\partial {\bf k}}\int V^{1,2}_{\bf k}\left({\bf k}_2-{\bf k}_1\right)\left[ ({\bf k}-{\bf k}_2)n_{{\bf k}}^2n_{2}^2\frac{\partial n_{1}}{\partial {\bf k}_1}
+({\bf k}_1-{\bf k})n_{{\bf k}}^2n_{1}^2\frac{\partial n_{2}}{\partial {\bf k}_2}
+({\bf k}_2-{\bf k}_1)n_{1}^2n_{2}^2\frac{\partial n_{{\bf k}}}{\partial {\bf k}} \right]\ {\rm d}{\bf k}_1 {\rm d} {\bf k}_2.
\end{align}
The bare expression for the interaction coefficient $V^{1,2}_{\bf k}$ follows from symmetrisation of Eq.~\eqref{eq:slam_localisation}:
\begin{align}\label{eq:V12k_bare}
 V^{1,2}_{\bf k} = V\left(\mathbf{k}_1, \mathbf{k}_2, \mathbf{k}\right)= \frac{24\pi}{|{\bf k}_2-{\bf k}_1|({\bf k}_2-{\bf k}_1)^2} f\left(\mathbf{k}, \mathbf{k}_2, \mathbf{k}_1\right) +  \frac{24\pi}{|\mathbf{k}_2-\mathbf{k}|(\mathbf{k}-\mathbf{k}_2)^2} \left[ f\left(\mathbf{k}_1, \mathbf{k}, \mathbf{k}_2\right)+ f\left(\mathbf{k}_1, \mathbf{k}_2, \mathbf{k}\right) \right].
\end{align}
However this expression leads to divergence of the integrand in the SLAM, and needs regularlisation. We carry this out in the next section.

Using \texttt{Mathematica}, we confirm that $V^{{\bf k}_1,{\bf k}_2}_{\bf k}$ has the symmetry properties $V^{1,2}_{\bf k}=V^{2,1}_{\bf k}=V^{1,{\bf k}}_2$, 
and has scale invariance $V^{\lambda{\bf k}_1,\lambda{\bf k}_2}_{\lambda{\bf k}} = |\lambda|^{-7}V^{{\bf k}_1,{\bf k}_2}_{\bf k}$.

\section{Regularisation of SLAM locality}

\noindent The interaction coefficient  $V^{1,2}_{\bf k}$ defined in Eq.~\eqref{eq:V12k_bare} contains a denominator that becomes zero when either or all $\mathbf{k}_1 =\mathbf{k}_2$, $\mathbf{k}_1 =\mathbf{k}$ or $\mathbf{k}_2 =\mathbf{k}$. Thus, for the SLAM to be well-posed, the integral in Eq.~\eqref{eq:slam-derivation} must converge at these singular points.

Direct examination of Eq.~\eqref{eq:V12k_bare}
shows that when $\epsilon=|\mathbf{k}_1 -\mathbf{k}| \to 0$ and $\mathbf{k}_2$ is fixed, $V^{1,2}_{\bf k}\propto \epsilon^0$ as $\epsilon\to 0$ and the square bracket is $\propto  \epsilon$ as $\epsilon \to 0$. Thus, the integral in Eq.~\eqref{eq:slam-derivation}
is convergent in this limit. By symmetry, it is also convergent when $\epsilon =|\mathbf{k}_2 -\mathbf{k}| \to 0$ and $\mathbf{k}_1$ is fixed. Similarly, when  $\epsilon= |\mathbf{k}_1 -\mathbf{k}_2| \to 0$ and $|\mathbf{k}_1 - \mathbf{k}|, |\mathbf{k}_2 - \mathbf{k}| = O(1)$, we have $V^{1,2}_{\bf k} \propto \epsilon^0$ as $\epsilon \to 0$ and the square bracket is $\propto  \epsilon$ as $\epsilon\to 0$. Thus again, the integral is convergent in this limit.

Now consider the limit when both $\mathbf{k}_1$ and $\mathbf{k}_2$ tend to $\mathbf{k}$ simultaneously. Introducing polar coordinates $\mathbf{k}_1 = {\bf k}+ r \cos( \theta)$ and $\mathbf{k}_2 = {\bf k} + r \sin(\theta)$,  we find $V^{1,2}_{\bf k} \propto r^{-7}$ as $r\to 0$ and the square bracket is $\propto r^{3}$ as $r\to 0$. Thus, the integral is divergent as $\int r^{-2}\, {\rm d} r$ in the limit $r \to 0$. In fact, the Taylor expansion involved in the derivation of the SLAM
fail when $\mathbf{k}_1$ and $\mathbf{k}_2$ tend to $\mathbf{k}$ simultaneously. We can track this failure to breaking the assumption $p,q \ll |\mathbf{k}_1-\mathbf{k}|, |\mathbf{k}_2-\mathbf{k}|$. Considering that the original WKE~\eqref{eq:kinetic} is convergent in this limit, we conclude that the respective SLAM divergence is spurious and should be eliminated by a regularization procedure.

For such a regularization, we borrow the idea of collision efficiency from the kinetic theory of sticky particles. This approach notes that the cross-section of a two-particle collision is significantly reduced when the particles have very separated sizes, and corrects the simple ``differential sedimentation'' model by introducing an effective cut-off (or a significant reduction) of the interaction of particles with separated sizes~\cite{horvai_coalescence_2008}.

Similarly, we will introduce a soft cut-off to tame the interactions when both
$\mathbf{k}_1$ and $\mathbf{k}_2$ are close to $\mathbf{k}$ simultaneously. Namely, our new interaction coefficient will be modified as follows,
\begin{align}\label{eq:VR}
    V^{1,2}_{\bf k} \to \frac{R^{1,2}_{\bf k}} {R^{1,2}_{\bf k}+M} V^{1,2}_{\bf k}, \quad
\text{where}\quad 
   R^{1,2}_{\bf k} =
   \frac{({\bf k}_1-{\bf k})^2}{{\bf k}_2^2} +
   \frac{({\bf k}_2-{\bf k})^2}{{\bf k}_1^2} +
   \frac{({\bf k}_1-{\bf k}_2)^2}{{\bf k}^2},   
\end{align}
and $M= {\rm constant} \ll 1$ is a locality parameter. After such a modification, the interaction coefficient retains all of its original scaling properties and symmetries with respect to exchanging indices (hence the wave energy conservation still holds, see the next section). At the same time, now $V^{1,2}_{\bf k} \propto r^{-5}$ as $r\to 0$ for both $\mathbf{k}_1$ and $\mathbf{k}_2$ tending to $\mathbf{k}$ simultaneously, so the resulting integral in SLAM is convergent.

The final expression for the regularised interaction coefficient of the SLAM is: 
\begin{align}
\label{eq:V12k}
 V^{1,2}_{\bf k}
=-&\left(\frac{R^{1,2}_{\bf k}} {R^{1,2}_{\bf k}+M}\right)
    \left( \frac{1296\pi^2}
        { ({\bf k} - {\bf k}_1)^3 ({\bf k} - {\bf k}_2)^3 ({\bf k}_1 - {\bf k}_2)^3  
          ({\bf k}   + {\bf k}_1 - 2{\bf k}_2)^3  
          ({\bf k}   + {\bf k}_2 - 2{\bf k}_1)^3  
          ({\bf k}_1 + {\bf k}_2 - 2{\bf k})^3 }
    \right) \Biggl[ \nonumber \\
&-\biggl\{  |{\bf k}-{\bf k}_1|  ({\bf k}-{\bf k}_1)^3  ({\bf k}+{\bf k}_1-2{\bf k}_2)^3
    \Bigl(
            4{\bf k}^4  -  18{\bf k}^3 {\bf k}_1 
            -  {\bf k}_2^2 ( 5{\bf k}^2 + 8{\bf k}{\bf k}_1 + 5{\bf k}_1^2 )\nonumber\\
    &\quad  +  2{\bf k}_2 ( {\bf k} + {\bf k}_1 ) ( {\bf k}^2 + {\bf k}{\bf k}_1 + {\bf k}_1^2 )
            +  25{\bf k}^2{\bf k}_1^2  -  18{\bf k}{\bf k}_1^3  +  6{\bf k}_2^3 ( {\bf k} + {\bf k}_1 ) 
            +  4{\bf k}_1^4  -  3{\bf k}_2^4
    \Bigr)
\biggr\}\nonumber\\ 
&+\biggl\{  |{\bf k}-{\bf k}_2|  ({\bf k}-{\bf k}_2)^3  ({\bf k}+{\bf k}_2-2{\bf k}_1)^3 
    \Bigl(
            4{\bf k}^4  -  18{\bf k}^3 {\bf k}_2 
            -  {\bf k}_1^2 ( 5{\bf k}^2 + 8{\bf k}{\bf k}_2 + 5{\bf k}_2^2 )\nonumber\\
    &\quad  +  2{\bf k}_1 ( {\bf k} + {\bf k}_2 ) ( {\bf k}^2 + {\bf k}{\bf k}_2 + {\bf k}_2^2 )
            +  25{\bf k}^2{\bf k}_2^2  -  18{\bf k}{\bf k}_2^3  +  6{\bf k}_1^3 ( {\bf k} + {\bf k}_2 ) 
            +  4{\bf k}_2^4  -  3{\bf k}_1^4
    \Bigr)
\biggr\}\nonumber \\
&-\biggl\{  |{\bf k}_1-{\bf k}_2|  ({\bf k}_1-{\bf k}_2)^3  ({\bf k}_1+{\bf k}_2-2{\bf k})^3 
    \Bigl(
            4{\bf k}_1^4  -  18{\bf k}_1^3 {\bf k}_2 
            -  {\bf k}^2 ( 5{\bf k}_1^2 + 8{\bf k}_1{\bf k}_2 + 5{\bf k}_2^2 )\nonumber\\
    &\quad  +  2{\bf k} ( {\bf k}_1 + {\bf k}_2 ) ( {\bf k}_1^2 + {\bf k}_1{\bf k}_2 + {\bf k}_2^2 )
            +  25{\bf k}_1^2{\bf k}_2^2  -  18{\bf k}_1{\bf k}_2^3  +  6{\bf k}^3 ( {\bf k}_1 + {\bf k}_2 ) 
            +  4{\bf k}_2^4  -  3{\bf k}^4
    \Bigr)
\biggr\}
\Bigg].
\end{align}

\section{Wave energy conservation by the 1D SLAM}

\noindent The 1D SLAM, Eq.~\eqref{eq:slam}, is a continuity equation for the wave action spectrum $n_{\bf k}$, so the wave action $N$ is conserved by definition. The wave energy $E$ can be expressed as an integral of the  wave energy spectrum $\omega_{\bf k} n_{\bf k}$ as defined in Eq.~\eqref{eq:invariants}. 
Hence, the time derivative of the wave energy can be expressed as 
\begin{align}\label{eq:energyeq}
\frac{\partial E}{\partial t}  &= \int \omega_{\bf k}\frac{\partial n_{\bf k}}{\partial t} \ {\rm d}{\bf k} = \left[ \omega_{\bf k}Q({\bf k})\right]_{-\infty}^{\infty} - \int  {\bf k}\, Q({\bf k}) \  {\rm d}{\bf k}\nonumber \\
&= - \frac{1}{\Lambda^{5/2}}\int  {\bf k}V^{1,2}_{\bf k}\left( {\bf k}_2- {\bf k}_1 \right)\left[ ({\bf k}-{\bf k}_2)n_{{\bf k}}^2n_{2}^2\frac{\partial n_{1}}{\partial {\bf k}_1}
+({\bf k}_1-{\bf k})n_{{\bf k}}^2n_{1}^2\frac{\partial n_{2}}{\partial {\bf k}_2}
+({\bf k}_2-{\bf k}_1)n_{1}^2n_{2}^2\frac{\partial n_{{\bf k}}}{\partial {\bf k}} \right]\ {\rm d}{\bf k}_1 {\rm d} {\bf k}_2 {\rm d} {\bf k},
\end{align}
where we have used integration by parts and the expression for $Q$ given in Eq.~\eqref{eq:slam}. Using the symmetry of the integral in~\eqref{eq:energyeq} with respect to swapping of the indices ${\bf k} \leftrightarrow {\bf k}_1$ and ${\bf k} \leftrightarrow {\bf k}_2$, as well as the properties of $V^{1,2}_{\bf k}$, we can re-express integral~\eqref{eq:energyeq} as
\begin{align}\label{eq:energytrans}
\frac{\partial E}{\partial t} = - \frac{1}{3\Lambda^{5/2}}\int  & \left\{ {\bf k}\left( {\bf k}_2- {\bf k}_1 \right) - {\bf k}_1({\bf k}_2 - {\bf k}) - {\bf k}_2({\bf k} - {\bf k}_1)\right\}V^{1,2}_{\bf k}\nonumber\\
&\times\left[ ({\bf k}-{\bf k}_2)n_{{\bf k}}^2n_{2}^2\frac{\partial n_{1}}{\partial {\bf k}_1}
+({\bf k}_1-{\bf k})n_{{\bf k}}^2n_{1}^2\frac{\partial n_{2}}{\partial {\bf k}_2}
+({\bf k}_2-{\bf k}_1)n_{1}^2n_{2}^2\frac{\partial n_{{\bf k}}}{\partial {\bf k}} \right]\ {\rm d}{\bf k}_1 {\rm d} {\bf k}_2 {\rm d} {\bf k}.
\end{align}
Integral~\eqref{eq:energytrans} vanishes due to the cancellations of the wavenumbers in the $\{\cdots\}$ brackets. This proves that the 1D SLAM~\eqref{eq:slam} also conserves the wave energy $E$.

\section{Convergence  of the integral $I(x)$}

\noindent We consider the convergence criteria of the collision integral in the 1D SLAM~\eqref{eq:slam} assuming a power-law profile for the wave action spectrum of the form $n_{\bf k} = Ck^x$. It is more convenient to study convergence using the dimensionless form, integral $I(x)$, from Eq.~\eqref{eq:I} in the main text. For convenience, we present it again below.
\begin{align}\label{eq:Isupp}
I(x) = -\frac{C^5}{\Lambda^{5/2}}\int x\, V^{{\bf s}_1,{\bf s}_2}_{1}\left({\bf s}_2-{\bf s}_1 \right)\left[ {\rm sgn}({\bf s}_1)(1-{\bf s}_2)s_2^{2x}s_1^{x-1} +{\rm sgn}({\bf s}_2)({\bf s}_1-1)s_1^{2x}s_2^{x-1} +({\bf s}_2-{\bf s}_1)s_1^{2x}s_2^{2x} \right] \, {\rm d}{\bf s}_1 {\rm d}{\bf s}_2.
\end{align}
Cases $x=0$ and $x=-2$ correspond to the limiting behaviour of the Rayleigh-Jeans spectrum. In these cases, the integrand of $I(x)$ is identically equal to zero, so the integral is trivially convergent. Thus, we consider $x\neq 0,-2$, and study the convergence of $I(x)$ in the cases when one variable tends to one of two limits,  $s_1\to 0$ or $s_1\to \infty$, while $s_2$ is kept as a nonzero constant. It is sufficient to study only the asymptotics of $s_1$ due to the symmetry with respect to ${\bf s}_1 \leftrightarrow {\bf s}_2$. Additionally, we also need to study the limit of both variables $s_1, s_2$ tending simultaneously to $0$ or $\infty$. We begin with the single variable limit.

\subsection{Limit $s_1 \ll s_2, 1$}

\noindent Using \texttt{Mathematica}, assuming $s_1 \ll s_2, 1$, and expanding $V^{{\bf s}_1,{\bf s}_2}_1$ in the limit of small $s_1$, we obtain in the leading order $V^{{\bf s}_1,{\bf s}_2}_1 \propto s_1^{0}$. Also, $ ({\bf s}_2-{\bf s}_1) \propto s_1^0$ and the square bracket $\Delta n:=\left[{\rm sgn}({\bf s}_1)(1-{\bf s}_2)s_2^{2x}s_1^{x-1} +{\rm sgn}({\bf s}_2)({\bf s}_1-1)s_1^{2x}s_2^{x-1} +({\bf s}_2-{\bf s}_1)s_1^{2x}s_2^{2x}\right]$ in Eq.~\eqref{eq:Isupp}, taking into account that $x\ne -2$, has the following leading order expression,
\begin{align*}
{\rm sgn}({\bf s}_1)(1-{\bf s}_2)s_2^{2x}s_1^{x-1} +(-{\rm sgn}({\bf s}_2)s_2^{x-1} +{\bf s}_2s_2^{2x}) s_1^{2x}.
\end{align*}
In this expression, the term proportional to $s_1^{x-1}$ dominates for $x>-1$, the term $\propto s_1^{2x}$ is dominant for $x<-1$, and both terms have comparable contributions for $x=-1$. Since the terms outside of the square brackets 
in Eq.~\eqref{eq:Isupp}
scale as $s_1^0$, we conclude that in the limit $s_1\to 0$,  with $s_2$ fixed, the integral $I(x)$ is convergent for $0<x$.

\subsection{Limit $s_1\gg s_2, 1$}

\noindent We now consider the opposite limit, in which the variable $s_1\to \infty$ and the other $s_2$ is fixed. Using \texttt{Mathematica}, we obtain in the leading order $V^{{\bf s}_1,{\bf s}_2}_1 \propto -s_1^{-7}{\rm sgn}({\bf s}_1)$, and we also have $({\bf s}_2-{\bf s}_1) \propto -{\bf s}_1$. 
We find that
\begin{align*}
\Delta n =    {\rm sgn}({\bf s}_1)(1-{\bf s}_2)s_2^{2x}s_1^{x-1} +{\rm sgn}({\bf s}_1) [{\rm sgn}({\bf s}_2)s_2^{x-1} +s_2^{2x}] s_1^{2x+1},
\end{align*}
where the term $\propto s_1^{x-1}$ dominates in the limit for $x<-2$ and the term $\propto s_1^{2x+1}$ dominates for $-2< x$. Hence, we conclude that in the limit of one large variable, $s_1\to \infty, \, s_2=~{\rm constant}$, the integral $I(x)$ is convergent for $x<2$.

\subsection{Two variables are small simultaneously, $s_1,s_2\to 0$}

\noindent Considering the limit when both ${\bf s}_1$ and ${\bf s}_2$ simultaneously go to zero, we first transform to polar coordinates
\begin{align*}
{\bf s}_1 = r\cos \left(\theta\right), \quad {\bf s}_2 = r\sin \left(\theta\right),
\end{align*}
so that
\begin{align*}
    \Delta n &= \left\{\left(1-r\, \sin \left(\theta\right)\right) \cos \left(\theta\right)  \left|\cos \left(\theta\right) \right|^{x-2} \left|\sin\left(\theta\right)\right|^{2x} +
    (1-r\, \cos \left(\theta\right)) \sin \left(\theta\right) \left|\sin\left(\theta\right)\right|^{x-2} \left|\cos \left(\theta\right)\right|^{2x}
    \right\} r^{3x-1}\\
    &+(\sin \left(\theta\right) - \cos \left(\theta\right)) \left|\cos \left(\theta\right) \sin \left(\theta\right) \right|^{2x}
    r^{4x+1},
\end{align*}
and consider the limit as $r\to 0$, giving
\begin{align*}
    \Delta n = \left\{ \cos \left(\theta\right) \, \left|\cos \left(\theta\right)\right|^{x-2} \left|\sin \left(\theta\right)\right|^{2x} +
     \sin \left(\theta\right) \, \left|\sin \left(\theta\right) \right|^{x-2} \left|\cos \left(\theta\right) \right|^{2x}
    \right\} r^{3x-1}
    +\left(\sin \left(\theta\right) - \cos \left(\theta\right)\right) \left|\cos\left(\theta\right) \sin \left(\theta\right) \right|^{2x}
    r^{4x+1}\, .
\end{align*}
Here, the terms $\propto r^{3x-1}$ and $\propto r^{4x+1}$ are dominant for $-2<x$ and $x<-2$ respectively. With \texttt{Mathematica}, we obtain the following asymptotic:
\begin{align}
V^{{\bf s}_1,{\bf s}_2}_1  \left({\bf s}_2 - {\bf s}_1\right)= V^{r \cos \left(\theta\right),r \sin \left(\theta\right)}_1 \left(r \left[\sin \left(\theta\right) - \cos \left(\theta\right)\right]\right) \underset{r \rightarrow 0}{\propto} {r}.
\end{align}
The change of integration variables from $s_1,s_2$ to $r,\theta$ yields the Jacobian $r$. Putting these asymptotic expressions together, we obtain that $I(x)$ is convergent for $r\to 0$ when $-2/3<x$.

\subsection{Two variables are large simultaneously, $s_1,s_2\to \infty$}

\noindent In this case, using the polar coordinates, we have
\begin{align*}
    \Delta n &= \left\{ \sin \left(\theta\right) \cos \left(\theta\right)  \left|\cos \left(\theta\right) \right|^{x-2} \left|\sin \left(\theta\right) \right|^{2x} - \cos \left(\theta\right) \sin \left(\theta\right) \left|\sin \left(\theta\right)\right|^{x-2} \left|\cos \left(\theta\right) \right|^{2x}
    \right\} r^{3x}\\
    &+\left(\sin \left(\theta\right) - \cos \left(\theta\right)\right) \left|\cos \left(\theta\right) \sin\left(\theta\right) \right|^{2x}
    r^{4x+1}.
\end{align*}
Here, the terms $\propto r^{3x}$ and $\propto r^{4x+1}$ are dominant for $x<-1$ and $x>-1$ respectively. Using \texttt{Mathematica}, we obtain the following asymptotic: 
\begin{align*}
V^{{\bf s}_1,{\bf s}_2}_1  \left({\bf s}_2 - {\bf s}_1\right)= V^{r \cos \left(\theta\right),r \sin \left(\theta\right)}_1 \left(r \left[\sin \left(\theta\right) - \cos \left(\theta\right)\right]\right)\propto {r}^{-6}.
\end{align*}
Putting these asymptotics together and taking into account the Jacobian $r$, we find that the integral $I(x)$ is convergent for $x<3/4$.

\subsection{Convergence of $I(x)$ --- summary}

 The above results on the convergence of $I(x)$ are summarised in the following table:

\begin{table}[h!]
\begin{tabular}{|c|c|}
\hline
\textbf{Limit} & \textbf{Region of Convergence}\\
\hline
$s_1 \to 0, s_2 = \hbox{const  (same for} \, s_2 \to 0, s_2 =
\hbox{const)}$ & $   x >0 $ \\ \hline
$s_1 \to \pm \infty, s_2 = \hbox{const  (same for} \, s_2 \to \pm \infty, s_2 =
\hbox{const)} $ & $x<  2$  \\ \hline
$ (s_1,s_2)=(r\cos \theta, r \sin \theta), \; \; r \to 0, \theta = \hbox{const}$ & $ -2/3 < x $ \\ \hline
$ (s_1,s_2)=(r\cos \theta, r \sin \theta), \; \; r \to \infty, \theta = \hbox{const}$ & $ x < 3/4 $ \\ \hline
\end{tabular}
\end{table}
Again, $I(x)$ is also convergent for the thermodynamic cases, $x=-2,0.$

\section{Proof of $I(2/5)=0$ via application of the Zakharov transform}

\noindent Let us split the integral $I(x)$ into three contributions defined by the three terms in the square bracket in Eq.~\eqref{eq:I}:  $I(x)=I_1(x)+I_2(x)+I_3(x)$ where
\begin{align}\label{eq:I_deconstructed}
I_1(x) &= -\frac{C^5x}{\Lambda^{5/2}}\int V^{{\bf s}_1,{\bf s}_2}_{1}\left({\bf s}_2-{\bf s}_1\right){\rm sgn}({\bf s}_1)(1-{\bf s}_2)s_2^{2x}s_1^{x-1} \ {\rm d}{\bf s}_1 {\rm d}{\bf s}_2,\nonumber\\
I_2(x) &= -\frac{C^5x}{\Lambda^{5/2}}\int V^{{\bf s}_1,{\bf s}_2}_{1}\left({\bf s}_2-{\bf s}_1\right){\rm sgn}({\bf s}_2)({\bf s}_1-1)s_1^{2x}s_2^{x-1} \ {\rm d}{\bf s}_1 {\rm d}{\bf s}_2,\nonumber\\
I_3(x) &= -\frac{C^5x}{\Lambda^{5/2}}\int V^{{\bf s}_1,{\bf s}_2}_{1}({\bf s}_2-{\bf s}_1)^2s_1^{2x}s_2^{2x} \ {\rm d}{\bf s}_1 {\rm d}{\bf s}_2.
\end{align}
Integral $I_2(x)$ is in fact identical to  $I_1(x)$, as seen by swapping the dummy integration variables ${\bf s}_1$ and ${\bf s}_2$.

We will make a transformation of integral $I_1(x)$ using the change of coordinates (Zakharov transform) ${\bf s}_1 = 1/\tilde{{\bf s}}_1$, ${\bf s}_2 =\tilde{{\bf s}}_2/\tilde{{\bf s}}_1$. The resulting expression becomes
\begin{align}\label{eq:I1}
I_1(x) = -\frac{C^5x}{\Lambda^{5/2}}\int V^{\tilde{\bf s}_1,\tilde{\bf s}_2}_{1}  \tilde{s}_1^{2x} \tilde{s}_2^{2x}   \tilde{\bf s}_1(\tilde{\bf s}_1-\tilde{\bf s}_2) (\tilde{\bf s}_2-1) \tilde{s}_1^{2-5x}  \ {\rm d}\tilde{\bf s}_1 {\rm d}\tilde{\bf s}_2,
\end{align}
where we have used scale invariance and the symmetry of $V^{{\bf s}_1,{\bf s}_2}_1$ to simplify $V^{1/\tilde{\bf s}_1,\tilde{\bf s}_2/\tilde{\bf s}_1}_1 = \tilde{s}_1^7V^{1,\tilde{\bf s}_2}_{\tilde{\bf s}_1} = \tilde{s}_1^7V^{\tilde{\bf s}_1,\tilde{\bf s}_2}_1$, and the fact that $\tilde{\bf s}_1 = {\rm sgn}(\tilde{\bf s}_1)\tilde{s}_1$.

Respectively, for integral $I_2(x)$ we write an expression by swapping the dummy integration variables ${\bf s}_1$ and ${\bf s}_2$ in $I_1(x)$, 
\begin{align}\label{eq:I2}
I_2(x) = -\frac{C^5x}{\Lambda^{5/2}}\int V^{\tilde{\bf s}_1,\tilde{\bf s}_2}_{1}  \tilde{s}_1^{2x} \tilde{s}_2^{2x}   \tilde{\bf s}_2(\tilde{\bf s}_2-\tilde{\bf s}_1) (\tilde{\bf s}_1-1) \tilde{s}_2^{2-5x}  \ {\rm d}\tilde{\bf s}_1 {\rm d}\tilde{\bf s}_2.
\end{align}


Setting $x=2/5$ and reconstructing the full integral $I(x)$ using the transformed expressions for $I_1$ and $I_2$ given by Eqs.~\eqref{eq:I1} and~\eqref{eq:I2} (with tildes dropped) respectively, together with the original expression for $I_{3}$ from~\eqref{eq:I_deconstructed} gives
\begin{align}
I(2/5) &= -\frac{2C^5}{5\Lambda^{5/2}}\int V^{{\bf s}_1,{\bf s}_2}_{1}s_1^{4/5}s_2^{4/5} \left[ {\bf s}_1 ({\bf s}_1-{\bf s}_2)({\bf s}_2-1)  + {\bf s}_2({\bf s}_2-{\bf s}_1)({\bf s}_1-1) + ({\bf s}_2- {\bf s}_1)^2\right] \ {\rm d}{\bf s}_1 {\rm d}{\bf s}_2.
\end{align}
From this, it is evident that the expression in the square bracket is zero. Hence we conclude that $I(2/5)=0$.

\section{Nonlocal evolution in the inverse cascade}

\noindent The convergence study of $I(x)$ indicates that the inverse wave action cascade KZ spectrum with power-law exponent $x=4/5$ leads to a  convergent WKE in the infrared region, but shows ultraviolet nonlocality. Namely, the collision integral of the WKE is divergent when both of the normalised integration wavenumbers $s_1$ and $s_2$ tend to infinity simultaneously.  This implies that the local theory cannot successfully describe the system, and indicates the need to derive a nonlocal model. The nature of the observed nonlocality of the inverse cascade KZ spectrum suggests that the evolution is dominated by interactions with the ultraviolet scales, namely when both $k_1$ and $k_2$ are much larger than $k$ in the integral of $Q_s(k)$. In this limit, the wave-action flux $Q_s(k)$ becomes 
\begin{align}
    Q_s(k) = -
    \frac{{\rm sgn}({\bf k})}{\Lambda^{5/2}}   \int_{k \ll k_{1},k_2}   V^{1,2}_{\bf k} \left( {\bf k}_2-{\bf k}_1\right)  \left[ -2{\bf k}_2n_{2}^2 n_{\bf k}^2\frac{ \partial n_{1}}{\partial {\bf k}_1}  + ({\bf k}_2-{\bf k}_1) n_{1}^2 n_{2}^2  \frac{\partial n_{\bf k}}{\partial {\bf k}}\right] {\rm d}{\bf k}_1 {\rm d}{\bf k}_2.\label{eq:Qnlanal1}
\end{align}

In the leading order, in the nonlocal limit $k \ll k_1, k_2$, we have $V^{1,2}_{\bf k} \approx V^{1,2}_0$, and the following symmetries hold,
\begin{align*}
  V^{1,2}_0 > 0,\quad V^{-1,-2}_0 = V^{1,2}_0,\quad \text{and}\quad V^{-1,2}_0 = V^{1,-2}_0.
\end{align*}

Because of these symmetries of $V^{1,2}_0$, the first term on the right hand side of Eq.~\eqref{eq:Qnlanal1} is zero (the contributions from the different quadrants of the $({\bf k}_1, {\bf k}_2)$-plane cancel each other). Hence,
\begin{align}\label{eq:Q-D}
    Q_s(k) = -D \frac{\partial n_{\bf k}}{\partial k}, \quad D =  \frac1{\Lambda^{5/2}}   \int_{k \ll k_{1},k_2} V^{1,2}_0 \, ({\bf k}_2 - {\bf k}_1)^2\,  n_{1}^2 n_{2}^2  \, {\rm d}{\bf k}_1 {\rm d}{\bf k}_2,
\end{align}
i.e., using the continuity equation for the wave action flux, we arrive at the heat equation,
\begin{align*}
      \frac{ \partial n_{\bf k}}{\partial t} = - \frac{\partial Q_s}{\partial k} = D \frac{\partial^2 n_{\bf k}}{\partial k^2}, 
\end{align*}
where the diffusion coefficient $D$ is positive because of the property $V^{1,2}_0 > 0$. This equation has a stationary scaling solution for the inverse wave action cascade, $Q_s(k)= {\rm const}<0$:
\begin{align*}
       n_{\bf k} =  \frac {|Q_s|}D k.
\end{align*}
The value of the exponent $x=1$ corresponds to the ultraviolet divergence of the collision integral $I(x)$ when both integration variables tend to infinity simultaneously. This makes our original ultraviolet nonlocality assumption consistent with the resulting spectrum because it confirms that the integral in $D$ is dominated by the ultraviolet scales, $k \ll k_{1}, k_2$. Of course, for convergence of this integral, there must be an effective ultraviolet cutoff, which can be naturally provided by the forcing scale $k_f$ and dissipation at a scale $k>k_f$.

\end{document}